\documentclass[letterpaper,journal]{IEEEtran}
\usepackage{amsmath,amssymb,amsfonts}

\usepackage{graphicx}
\usepackage[font=footnotesize]{caption}
\usepackage{subcaption}
\usepackage[inkscapelatex=false]{svg}
\usepackage{pdflscape}
\usepackage{siunitx}

\usepackage{array}
\usepackage{booktabs}
\usepackage{multirow}
\usepackage{makecell}
\usepackage{tabularx}

\usepackage{algorithm}
\usepackage{algpseudocode}

\usepackage{hyperref}
\usepackage{url}
\usepackage{verbatim}
\usepackage{stfloats}
\usepackage{textcomp}
\usepackage[nolist]{acronym}
\usepackage{pdfpages}
\usepackage{tocloft}
\usepackage{tikz}
\usepackage{color}
\usepackage{cite}

\def\BibTeX{{\rm B\kern-.05em{\sc i\kern-.025em b}\kern-.08em
    T\kern-.1667em\lower.7ex\hbox{E}\kern-.125emX}}
\usepackage{balance}

\begin{document}

\let\codexOldAcrodef\acrodef
\let\acrodef\newacro
\acrodef{DSP}{digital signal processing}
\acrodef{ASIC}{application-specific integrated circuit}
\acrodef{IM/DD}{intensity modulation and direct detection}
\acrodef{LO}{local oscillator}
\acrodef{CFO}{carrier frequency offset}
\acrodef{EVM}{error vector magnitude}
\acrodef{16QAM}{16-quadrature amplitude modulation}
\acrodef{QAM}{quadrature amplitude modulation}
\acrodef{RC}{raised cosine}
\acrodef{FLL}{frequency-locked loop}
\acrodef{PLL}{phase-locked loop}
\acrodef{AFC}{automatic frequency control}
\acrodef{STO}{short-time autocorrelation}
\acrodef{NCO}{numerically controlled oscillator}
\acrodef{NDA}{non-data-aided}
\acrodef{LEO}{low Earth orbit}
\acrodef{FFT}{fast Fourier transform}
\acrodef{RF}{radio-frequency}
\acrodef{DDCPR}{decision-directed carrier phase recovery}
\acrodef{IIR}{infinite impulse response}
\acrodef{BER}{bit error ratio}
\acrodef{FEC}{forward error correction}
\acrodef{HTF-FOC}{hybrid time-frequency domain frequency offset compensation}
\acrodef{FOC}{frequency offset compensation}
\acrodef{LiFi}{light fidelity}
\acrodef{QPSK}{quadrature phase shift keying}
\acrodef{VV}{Viterbi-Viterbi}
\acrodef{RRC}{root raised cosine}
\acrodef{DAC}{digital-to-analog converter}
\acrodef{IQ}{in-phase and quadrature}
\acrodef{CW}{continuous wave}
\acrodef{LED}{light-emitting diode}
\acrodef{FSO}{free-space optical}
\acrodef{BPSK}{binary phase shift keying}
\acrodef{SMF}{single mode fiber}
\acrodef{AWGN}{additive white Gaussian noise}
\acrodef{ADC}{analog-to-digital converter}
\acrodef{SPS}{samples per symbol}
\acrodef{ISI}{inter-symbol interference}
\acrodef{PSD}{power spectral density}
\acrodef{CMA}{constant modulus algorithm}
\acrodef{OIPLL}{optical injection locking}
\acrodef{NTN}{non-terrestrial network}
\acrodef{SNR}{signal-to-noise ratio}
\acrodef{MZM}{Mach-Zehnder modulator}
\acrodef{ML}{maximum likelihood}
\acrodef{PI}{proportional-integral}
\acrodef{DCO}{digital coherent optics}
\acrodef{OWC}{optical wireless communication}
\acrodef{WDM}{wavelength division multiplexing}
\acrodef{QSFP-DD}{quad small form-factor pluggable double density}
\acrodef{OSFP}{octal small form-factor pluggable}
\acrodef{TIA}{transimpedance amplifier}
\acrodef{DD}{decision-directed}
\acrodef{PAT}{pointing, acquisition, and tracking}
\acrodef{SER}{symbol error rate}
\acrodef{OGR}{optical ground receiver}
\acrodef{PDF}{probability density function}
\acrodef{SPE}{sampling phase error}
\acrodef{AOA}{angle-of-arrival}
\acrodef{SFE}{sampling frequency error}
\acrodef{PEP}{pairwise error probability}
\acrodef{5G}{fifth generation}
\acrodef{6G}{sixth generation}
\acrodef{NASA}{national aeronautics and space administration}
\acrodef{TB}{terabyte}
\acrodef{TEM}{transverse electromagnetic}
\acrodef{RMS}{root mean square}
\acrodef{i.i.d.}{independent and identically distributed}
\acrodef{PD}{photodiode}
\acrodef{DC}{direct current}
\acrodef{CPR}{carrier phase recovery}
\acrodef{OSR}{oversampling ratio}
\acrodef{LPF}{low-pass filter}

\let\acrodef\codexOldAcrodef

\title{Hybrid Time-Frequency Domain Frequency Offset Compensation Under GHz Doppler Shift for LEO Satellite-to-Ground\\ Coherent Free-Space Optical Communication}
\author{Tiankuo Jiao, Hossein Kazemi, and Harald Haas
\thanks{This work was supported by the Engineering and Physical Sciences Research Council (EPSRC) Centre for Doctoral Training (CDT) in Photonic and Electronic Systems (PES) and the EPSRC project `Future Communications Hub in All-Spectrum Connectivity (HASC)' under grant EP/Y037197/1. The authors are with the LiFi Research and Development Center, Electrical Engineering Division, Department of Engineering, University of Cambridge, CB3 0FA Cambridge, United Kingdom (email: tj361@cam.ac.uk).}\vspace{-15pt}}


\maketitle

\begin{abstract}
Coherent free-space optical (FSO) communication is a promising solution for low Earth orbit (LEO) satellite downlink transmission. However, high orbital velocity introduces multi-GHz Doppler shifts that appear as a rapidly time-varying carrier frequency offset (CFO), which is a major challenge for conventional coherent optical receivers. Narrow-range digital loops cannot acquire the initial offset, while wide-range feedforward or optical domain solutions either leave large residual errors or impose substantial implementation cost. In this paper, a Doppler-aware hybrid time-frequency frequency offset compensation (HTF-FOC) receiver architecture is proposed for coherent LEO satellite-to-ground FSO links using a cumulative time-varying random process model for the dynamic Doppler-induced phase shift. The proposed receiver implements a hybrid acquisition and tracking procedure to acquire and compensate for multi-GHz Doppler variations, including 4th-power FFT-based coarse CFO acquisition, residual CFO handover verification, and low-complexity decision-directed (DD) frequency-locked loop (FLL) tracking. The phase-averaged pairwise error probability (PEP) and union-bound symbol error rate (SER) expressions are derived and verified using Monte Carlo simulations. The results demonstrate that the proposed HTF-FOC method tracks Doppler shifts beyond \textpm5~GHz while keeping the residual CFO below 80~MHz with a success rate of 100\% for typical LEO altitudes of 400--800~km and orbital speeds of 7.3--7.9~km/s.
\end{abstract}


\section{Introduction}
The deployment of \ac{LEO} satellite mega-constellations for broadband access \cite{osoro2025emissions}, Earth observation, and \acp{NTN} in future \ac{6G} systems is increasing the demand for high-capacity space-to-ground links \cite{aerospace12060550, mahboob2024revolutionizing, shayea2024integration}. Although \ac{RF} systems remain central to global connectivity, spectrum scarcity and interference motivate the use of optical technologies \cite{tarhouni2025free, kaushal2016optical, xu2025ground}. In particular, coherent \ac{FSO} communication is an attractive choice since it preserves phase information, supports high spectral efficiency, and can naturally interface with the fiber-based \ac{WDM} infrastructure. In an all-optical front-end, the transmitted optical signal is launched from fiber to free space and is recoupled back into a fiber at the receiver, without intermediate optical-to-electrical conversion at the terminal. This architecture preserves optical bandwidth, supports remote coherent \ac{FSO} transceiver placement \cite{lorences2020200}, and is compatible with commercial \ac{DCO} modules such as 400~Gb/s \ac{OSFP} and \ac{QSFP-DD} transceivers in size-, weight-, and power-constrained satellite \ac{FSO} terminals \cite{10138358}. Experimental demonstrations of 100~Gb/s-class and higher coherent \ac{FSO} links showcase the feasibility of high-throughput optical downlink \cite{walsh2022demonstration}.

In a coherent \ac{FSO} system, a narrow-linewidth \ac{CW} laser is amplitude- and phase-modulated by using a dual-parallel \ac{MZM}. The received optical field is mixed with a narrow-linewidth \ac{LO} to recover both the amplitude and phase. Unlike \ac{IM/DD} \ac{FSO}, which detects only intensity, coherent detection preserves spectral efficiency \cite{tan2021coherent}. Coherent receivers also provide a \ac{DSP} platform for mitigating carrier drift \cite{fernandes2022mitigation, almonacil2020digital,pita2022all,xu2025robust}, phase noise \cite{xu2025robust}, and \ac{IQ} impairments \cite{RecentAdvances}. However, \ac{LEO} satellite-to-ground coherent \ac{FSO} links introduce a dynamical impairment that is much more severe than in terrestrial fiber systems. With orbital speeds around 7.5~km/s, the optical carrier experiences Doppler shifts on the order of several GHz \cite{10138358, shoji2012pilot, mosnier2024free}. As the satellite traverses an orbit, the Doppler component changes sign and magnitude, producing a large, time-varying \ac{CFO}. Without compensation, the accumulated phase rotation distorts the constellation and substantially degrades symbol decisions.

Existing \ac{CFO} estimators for coherent optical detection exhibit a three-way trade-off among acquisition range, tracking capability, and implementation cost. Conventional digital \ac{PLL} and \ac{FLL} structures are effective for small residual offsets but cannot directly acquire rapidly time-varying multi-GHz \acp{CFO} \cite{coherbook,freq1,paillier2020space}. Feedforward \ac{STO} can capture offsets up to 4.5~GHz, but reported residual errors remain on the order of hundreds of MHz \cite{fan2024foe}. The $M$th-power \ac{VV} estimator provides wide-range coarse acquisition, though its accuracy is limited by \ac{FFT} resolution \cite{viterbi1983nonlinear,li2019phase}. Recent real-time \ac{DSP}-assisted \ac{AFC} experiments have demonstrated $\pm8$~GHz Doppler frequency shift compensation at 2.5~Gbaud \ac{QPSK}, showing the practical relevance of wide-range digital acquisition \cite{tang2024afc}. \Ac{ML}, cyclostationary, and Kalman-based approaches can improve accuracy \cite{meiyappan2012complex, bolcskei2002blind, xu2025robust}, but their computational burden makes them less attractive for real-time \ac{LEO} receivers. In addition, transmitter-side pre-compensation relies on accurate Doppler prediction \cite{almonacil2020digital}, and \ac{OIPLL} solutions extend the lock range at the cost of additional hardware for optical processing \cite{shoji2012pilot}. These shortcomings motivate a Doppler-aware coarse-to-fine \ac{DSP} receiver for \ac{LEO} coherent \ac{FSO} links.

Table~\ref{compute} provides a comparison between the aforementioned \ac{FOC} techniques \cite{fan2024foe,zhao2015digital,li2019phase,meiyappan2012complex,bolcskei2002blind,xu2025robust,REN2018166,almonacil2020digital,shoji2012pilot,tang2024afc,coherbook,freq1,paillier2020space}. In this paper, a \ac{HTF-FOC} method is systematically developed for coarse-to-fine acquisition and tracking of Doppler-induced multi-GHz \ac{CFO} trajectories in \ac{LEO} satellite-to-ground coherent \ac{FSO} links, which is a major challenge. The proposed method performs 4th-power \ac{FFT}-based coarse acquisition, verifies the residual phase increment to activate fine tracking, then tracks the residual \ac{CFO} using a time-domain \ac{DD}-\ac{FLL}, without the need for optical locking, ephemeris pre-compensation, or adaptive equalization. In a previous work \cite{TJiao2026ICC}, the authors presented preliminary results for \ac{HTF-FOC} in a short coherent \ac{OWC} link, where mismatches between \ac{LO} lasers of the transmitter and receiver cause multiple hundreds of MHz \ac{CFO}. The main contributions of this paper are summarized as follows:

\begin{table*}[htbp]
\centering
\caption{Comparison of \ac{CFO} Estimation and Compensation Methods for Coherent \ac{FSO} Links}
\label{compute}
\begin{tabular}{lccc}
\toprule
\textbf{Method}

& \textbf{Capture Range} 
& \textbf{Tracking Ability} 
& \textbf{Complexity} \\
\midrule

\ac{STO} \cite{fan2024foe}
& Very large ($\pm 4.5$ GHz)
& Poor (static only)
& Low (simple correlation) \\

Pilot-based \ac{CFO} \cite{zhao2015digital} 
& Medium
& Moderate
& Low (per-symbol phase ops) \\

$M$-th power (Viterbi--Viterbi)  \cite{li2019phase}
& Large (after $4\times$ folding)
& None (coarse only)
& Moderate (one large \ac{FFT}) \\

\ac{ML}-based \ac{CFO} \cite{meiyappan2012complex}
& Very large
& None (batch only)
& High (grid search) \\

Cyclostationary estimator \cite{bolcskei2002blind}
& Large
& None (batch only)
& High (cyclic statistics) \\

Kalman / Extended Kalman \cite{xu2025robust}
& Medium
& Excellent
& High (matrix updates) \\

Blind \ac{CMA} \ac{CFO} \cite{REN2018166}
& Small
& Weak
& Moderate (iterative update) \\

Digital pre-compensation \cite{almonacil2020digital}
& Very large
& None
& Low (single rotation) \\

Optical \ac{PLL} / \ac{OIPLL} \cite{shoji2012pilot}
& Very large ($\pm 2.5$ GHz)
& Good
& N/A (optical-domain) \\

DSP-assisted \ac{AFC} \cite{tang2024afc}
& Very large ($\pm 8$ GHz)
& Good (real-time \ac{QPSK})
& Moderate--high (FPGA \ac{AFC}) \\

Conventional digital \ac{PLL} / \ac{FLL} \cite{coherbook,freq1,paillier2020space}
& Small ($\sim$MHz-level)
& Good (slow dynamics)
& Low (scalar loop ops) \\

\textbf{Proposed \ac{HTF-FOC}}
& \textbf{Multi-GHz}
& \textbf{Strong (per-symbol tracking)}
& \textbf{Moderate (one \ac{FFT} + light loop)} \\
\bottomrule
\end{tabular}
\end{table*}

\begin{figure*}[htbp]
    \centering
    \includegraphics[width=0.9\textwidth]{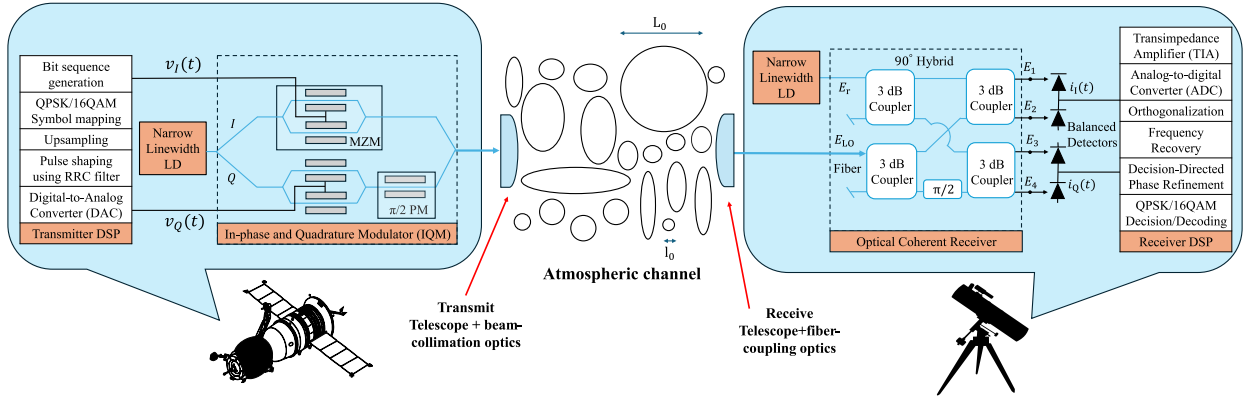}
    \caption{End-to-end architecture of the \ac{LEO} satellite-to-ground coherent \ac{FSO} communication system.}
    \label{fig:system}
\end{figure*}

\begin{itemize}
  \item The end-to-end coherent transmitter-to-receiver system, including optical \ac{IQ} modulation, free-space propagation, balanced photodetection, frequency recovery, and \ac{DD} phase refinement, is established for both \ac{QPSK} and \ac{16QAM}.
  \item The satellite-to-ground \ac{FSO} channel model is formulated by combining beam propagation, \ac{LEO} orbital geometry-based Doppler evolution, and turbulence-induced fading. Instead of a static frequency offset, the dynamic Doppler phase shift is incorporated using a time-varying process.
  \item A hybrid \ac{HTF-FOC} receiver architecture is developed for multi-GHz \ac{CFO} capture and per-symbol residual Doppler frequency tracking, with complexity analysis confirming a moderate computational cost for each algorithm.
  \item A \ac{CFO} handover criterion is proposed to verify the post-acquisition residual \ac{CFO} to ensure stable closed-loop fine tracking based on \ac{DD}-\ac{FLL}.
  \item The average \ac{PEP} and \ac{SER} are derived under the residual \ac{CFO}, and validated by Monte Carlo simulations. They are used to evaluate the impact of the residual \ac{CFO} for various \ac{LEO} link configurations and atmospheric turbulence scenarios.
\end{itemize}

The remainder of the paper is organized as follows. The end-to-end system model of the \ac{LEO} satellite-to-ground coherent \ac{FSO} link is established in Section~\ref{sec:system}. The proposed \ac{HTF-FOC} method is expounded in Section~\ref{sec:method}. Performance analysis in terms of \ac{PEP} and \ac{SER} in the presence of the residual \ac{CFO} is presented in Section~\ref{sec:pep}. Simulation results are discussed in Section~\ref{sec:results}. The concluding remarks are given in Section~\ref{sec:conclusions}.

\section{End-to-End System Modeling} \label{sec:system}
To elaborate, Fig.~\ref{fig:system} depicts the geometric configuration of the \ac{LEO} satellite-to-ground coherent \ac{FSO} link. In the following, coherent optical transmission, \ac{LEO} satellite-to-Earth \ac{FSO} channel, and coherent optical reception are established.

\newcounter{mycounter1}
\setcounter{mycounter1}{\value{equation}}
\setcounter{equation}{6}
\begin{figure*}[htbp]
\begin{equation}
E_{\mathrm{r}}(t) = \sqrt{\eta_{\mathrm{tx}}\eta_{\mathrm{rx}}P_{\mathrm{t}}L_{\mathrm{geo}}(R)\,L_{\mathrm{atm}}(R)h_{\mathrm{pe}}(t)I_\mathrm{r}(t)}\,
x(t-\tau)\,\exp\Big\{j\big[2\pi f_\mathrm{c}t+\phi_{\mathrm{D}}(t)+\phi_{\mathrm{turb}}(t)+\phi_{\mathrm{tx}}(t)\big]\Big\}
+z(t).
\label{eq:fso_passband_complex}
\end{equation}
\hrulefill
\end{figure*}
\setcounter{equation}{\value{mycounter1}}

\subsection{Coherent Optical Transmitter}
In a coherent \ac{FSO} transmitter, the optical signal is generated by driving the two arms of an \ac{IQ} modulator using the input data stream and the laser beam is launched toward the ground terminal through a telescope with beam collimation optics. A standard \ac{IQ} modulator composed of two parallel \ac{MZM} arms with a relative phase shift of $\pi/2$ is used to modulate the optical carrier \cite{11016179}. The transfer characteristic of each \ac{MZM} is approximately a cosine function \cite{seimetz2009high}. The input optical field $E_{\mathrm{in}}(t)=E_{0}\exp\!\left\{j\left[2\pi f_{\mathrm{c}}t+\phi_{\mathrm{tx}}(t)\right]\right\}$ is provided by a \ac{CW} laser with amplitude $E_{0}$, carrier frequency $f_{\mathrm{c}}$, and phase noise $\phi_{\mathrm{tx}}(t)$ due to the finite linewidth of the transmitter laser source. For a laser wavelength $\lambda$, $f_\mathrm{c}=c/\lambda$, with $c$ denoting the speed of light. The output optical field of the \ac{IQ} modulator is \cite{seimetz2009high}:
\begin{equation}
\begin{aligned}
E_{\mathrm{out}}(t) &= A_{\mathrm{IQ}}(t) E_{\mathrm{in}}(t), \\
&= A_{\mathrm{IQ}}(t) E_{0}\exp\!\left\{j\left[2\pi f_{\mathrm{c}}t+\phi_{\mathrm{tx}}(t)\right]\right\},
\end{aligned}
\label{eq:eout}
\end{equation}
where $A_{\mathrm{IQ}}(t)$ represents the field transfer function \cite{seimetz2009high}:
\begin{equation}
A_{\mathrm{IQ}}(t) = \frac{1}{2}\left[\cos\!\left(\frac{\pi V_{\mathrm{I}}(t)}{2V_{\pi}}\right)
+j\cos\!\left(\frac{\pi V_{\mathrm{Q}}(t)}{2V_{\pi}}\right)\right],
\label{eq:stx_def}
\end{equation}
where $V_{\pi}$ denotes the half-wave voltage, and $V_{\mathrm{I}}(t)$ and $V_{\mathrm{Q}}(t)$ are the driving voltages of the \ac{IQ} modulator, given by:
\begin{equation}
V_{\mathrm{I}}(t) = \frac{4V_{\pi}}{\pi}\zeta x_{\mathrm{I}}(t)+V_{\mathrm{DC}}, \,\,\,
V_{\mathrm{Q}}(t) = \frac{4V_{\pi}}{\pi}\zeta x_{\mathrm{Q}}(t)+V_{\mathrm{DC}},
\label{eq:IQ_Voltages}
\end{equation}
where $x_{\mathrm{I}}(t)=\Re\{x(t)\}$, $x_{\mathrm{Q}}(t)=\Im\{x(t)\}$, $x(t)$ denotes the baseband modulated signal, with $\Re$ and $\Im$ representing the real part and the imaginary part of a complex variable, respectively, $V_{\mathrm{DC}}$ is the \ac{DC} bias voltage, and $\zeta$ is a scaling factor to ensure that the signal variations in each arm remain within the dynamic range of the \ac{IQ} modulator. By operating both arms of the \ac{IQ} modulator at the quadrature point with a \ac{DC} bias of $V_{\mathrm{DC}}=-V_{\pi}/2$ and under small-signal modulation such that $|\zeta x_{\mathrm{I}}(t)|\ll1$ and $|\zeta x_{\mathrm{Q}}(t)|\ll1$, the nonlinear cosine transfer in \eqref{eq:stx_def} is linearized, and consequently:
\begin{equation}
A_{\mathrm{IQ}}(t) \approx A_{0} + \zeta x_{\mathrm{I}}(t)+j \zeta x_{\mathrm{Q}}(t) = A_{0} + \zeta x(t),
\label{eq:stx_approx}
\end{equation}
where $A_{0}=\frac{1}{2}(1+j)$. Hence, the varying part of the passband optical field in \eqref{eq:eout} is:
\begin{equation}
\tilde{E}_{\mathrm{out}}(t) \propto x(t)\exp\!\left\{j\left[2\pi f_{\mathrm{c}}t+\phi_{\mathrm{c}}(t)\right]\right\}.
\end{equation}
Furthermore, the continuous-time complex baseband signal is: 
\begin{equation}
x(t)=\sum_{n=0}^{\infty} x[n]\,g_{\mathrm{RRC}}(t-nT_{\mathrm{s}}),
\label{eq:sbb}
\end{equation}
where $x[n]=x_{\mathrm{I}}[n]+j\,x_{\mathrm{Q}}[n]$ is the complex symbol drawn from the signal constellation, $g_{\mathrm{RRC}}(t)$ is the impulse response of the pulse shaping filter, and $T_{\mathrm{s}}$ is the symbol period. To constrain the occupied spectral bandwidth, a \ac{RRC} filter with a roll-off factor of $0.1$ is used \cite{Proakis2008DigitalCommunications}. Without loss of generality, two modulation schemes of \ac{QPSK} and \ac{16QAM} are considered. For \ac{QPSK}, the input bit stream $b[n]$ is grouped into two-bit symbols such that $x_{\mathrm{I}}[n]=1-2b[2n]$ and $x_{\mathrm{Q}}[n]=1-2b[2n+1]$. For \ac{16QAM}, four bits form a symbol using $x_{\mathrm{I}}[n],x_{\mathrm{Q}}[n]\in\{\pm1,\pm3\}$ according to Gray mapping.

\begin{figure}[t!]
\centering
\includegraphics[width=0.7\linewidth]{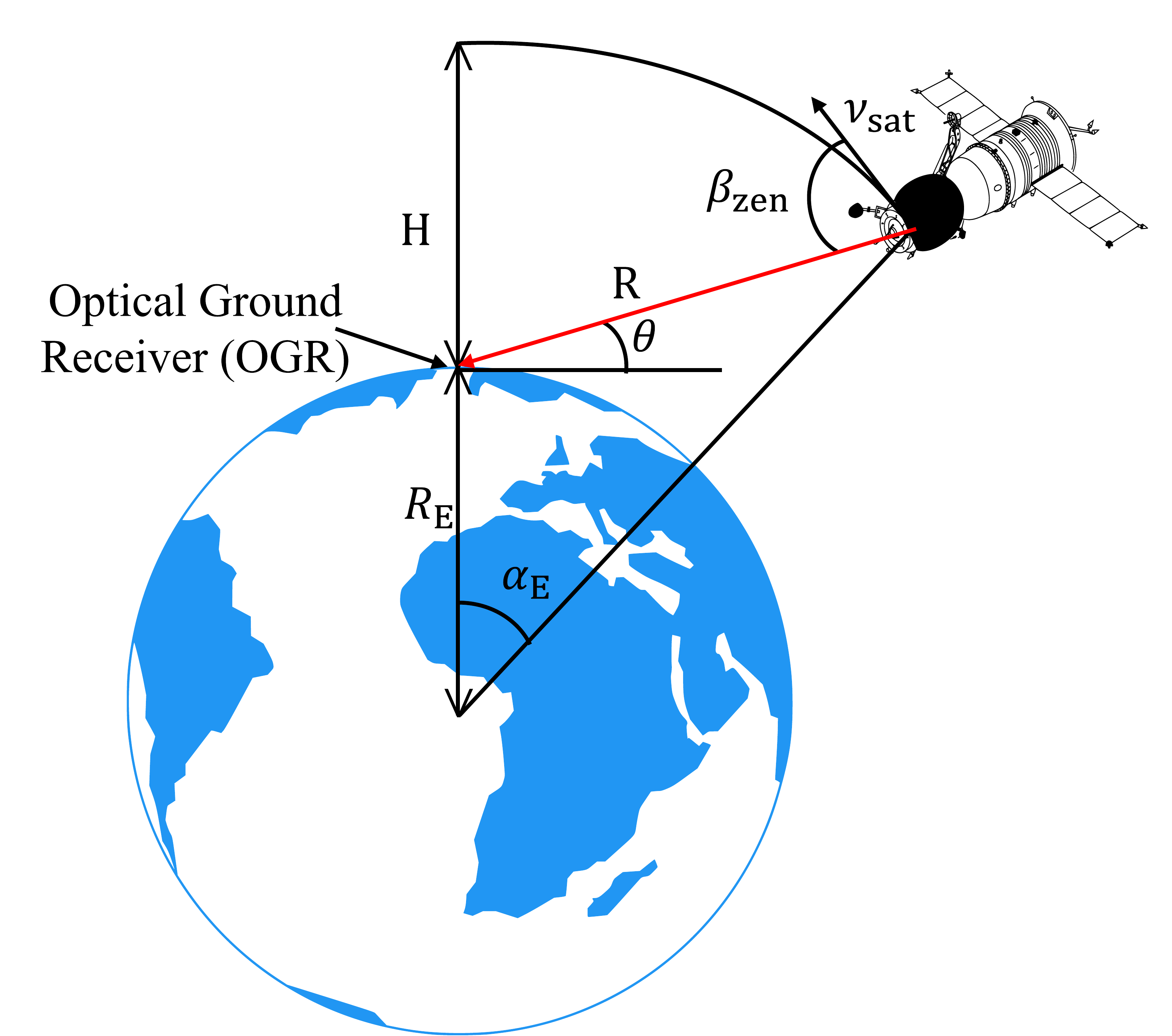}
\caption{Geometry of the \ac{LEO} satellite-to-ground link and observation angles. Note that all the vectors shown here are coplanar and the Earth-satellite radial direction vector $\mathbf{R}$ is perpendicular to the satellite velocity vector $\mathbf{v}_{\mathrm{sat}}$.\vspace{-10pt}}
\label{fig:satellite_geometry}
\end{figure}

\subsection{Satellite-to-Earth Channel Model}
The satellite-to-Earth \ac{FSO} channel is modeled by taking into account the effects of the Doppler shift $\phi_{\mathrm{D}}(t)$, geometric loss $L_{\mathrm{geo}}(R)$ and atmospheric attenuation $L_{\mathrm{atm}}(R)$ over the slant range $R$, pointing error loss $h_{\mathrm{pe}}(t)$, and turbulence-induced fading with instantaneous irradiance $I_\mathrm{r}(t)$ and phase $\phi_{\mathrm{turb}}(t)$. The received passband optical field is given in \eqref{eq:fso_passband_complex}, presented at the top of the page, where $P_{\mathrm{t}}$ is the transmitted optical power, $\eta_{\mathrm{tx}}$ and $\eta_{\mathrm{rx}}$ are the transmitter and receiver efficiencies, respectively, $\tau$ is the propagation delay, and $z(t)$ is the additive noise.

\subsubsection{Doppler Effect, Geometric Loss, and Pointing Error}
The Doppler effect arises from the relative motion of the \ac{LEO} satellite with respect to the \ac{OGR} \cite{10138358, 10347224}. It produces a time-varying phase shift that appears in the received signal as:
\addtocounter{equation}{1}
\begin{equation}
    \phi_{\mathrm{D}}(t) = 2\pi\int_{0}^{t}\!\big(f_{\mathrm{D}}(\xi)+f_{\mathrm{l}}\big)\,\mathrm{d}\xi,
    \label{eq:doppler_phase_shift}
\end{equation}
where $f_{\mathrm{D}}(t)$ is the Doppler-induced \ac{CFO}, and $f_{\mathrm{l}}$ accounts for the intrinsic laser frequency offset. This causes a continuous rotation of the received signal constellation \cite{10138358}. Fig.~\ref{fig:satellite_geometry} depicts the geometry of an \ac{LEO} satellite-to-ground link with these parameters: the elevation angle $\theta$ observed at the ground station, the Earth radius $R_{\mathrm{E}}$, the orbital altitude $H$, the satellite angular position $\beta_{\mathrm{zen}}$ relative to zenith, the satellite orbital velocity $v_{\mathrm{sat}}$, and the Earth central angle $\alpha_{\mathrm{E}}$. The slant range $R$ satisfies $R^{2}=R_{\mathrm{E}}^{2}+(R_{\mathrm{E}}+H)^{2}-2R_{\mathrm{E}}(R_{\mathrm{E}}+H)\cos\alpha_{\mathrm{E}}$. The elapsed pass time, which characterizes the temporal evolution of a satellite pass when viewed from the ground, is given by $\tau_{\mathrm{ela}}=\alpha_{\mathrm{E}}\sqrt{(R_{\mathrm{E}}+H)^3/\mu_E}$ \cite{shoji2012pilot}. The Doppler frequency shift can be expressed as \cite{10138358}:
\begin{equation}
f_{\mathrm{D}} = f_{\mathrm{c}}
\left(
\frac{
\sqrt{1 - \frac{v_{\mathrm{sat}}^{2}}{c^{2}}} - \left(1-\frac{v_{\mathrm{sat}}}{c}\cos\beta_{\mathrm{zen}}\right)
}{
1-\frac{v_{\mathrm{sat}}}{c}\cos\beta_{\mathrm{zen}}
}
\right),
\label{eq:doppler_shift_new}
\end{equation}
where $v_{\mathrm{sat}}=\sqrt{\mu_E/(R_{\mathrm{E}}+H)}$ from circular-orbit mechanics, and $\mu_E$ is the standard gravitational parameter of Earth \cite{10347224}.

In a coherent \ac{FSO} system with a fiber-to-free-space launch, the output beam is approximated by the lowest-order \ac{TEM} mode (i.e., TEM$_{00}$), which is a Gaussian beam \cite{Saleh2019Optics}. Hence, the intensity at a radial distance $\rho$ on the transverse plane located at distance $R$ from the beam waist is \cite{Saleh2019Optics}:
\begin{equation}
I(\rho,R)=\frac{2P_{\mathrm{t}}}{\pi W^2(R)}
\exp\!\left(-\frac{2\rho^2}{W^2(R)}\right),
\label{eq:gaussian_irradiance_R}
\end{equation}
where $W(R)=W_0\sqrt{1+\left(R/(\pi W_0^2/\lambda )\right)^2}$ is the radius of the $1/e^2$ beam spot, and $W_0$ is the radius of the beam waist. For a coaxially aligned circular aperture of radius $a$ at the receiver, the collected power is
$P_{\mathrm{r}}(R)=\int_0^{a} I(\rho,R)\,2\pi\rho\,\mathrm{d}\rho$, which leads to the geometric
spreading loss:
\begin{equation}
L_{\mathrm{geo}}(R) = \frac{P_{\mathrm{r}}(R)}{P_{\mathrm{t}}}
=1-\exp\!\left(-\frac{2a^2}{W^2(R)}\right).
\label{eq:Lgeo_R}
\end{equation}
In addition, following the Beer-Lambert law, the atmospheric attenuation due to absorption and scattering over a propagation distance $R$ is given by \cite{AndrewsPhillips2005}:
\begin{equation}
L_{\mathrm{atm}}(R)=\exp(-\kappa R),
\label{eq:Latm_general}
\end{equation}
where $\kappa$ denotes the altitude-dependent extinction coefficient averaged over the \ac{LEO} slant path.

In satellite-to-ground \ac{FSO} links, pointing jitter is a critical challenge. The jitter components of the received optical beam along the horizontal and vertical axes on the transverse plane are modeled as \ac{i.i.d.} zero-mean Gaussian random variables with a variance $\sigma_{\mathrm{pe}}^2$, where $\sigma_{\mathrm{pe}}$ is the \ac{RMS} jitter. Under a small-angle approximation, the equivalent radial displacement on the receiver plane, denoted by $\rho_{\mathrm{pe}}(t)$, follows a Rayleigh distribution \cite{farid2007outage}. For a Gaussian beam, the attenuation caused by pointing jitter can be expressed as \cite{farid2007outage}:
\begin{equation}
h_{\mathrm{pe}}(t)=h_{0}\exp\!\left(-\frac{\rho_{\mathrm{pe}}^2(t)}{W^{2}(R_{\mathrm{eq}})}\right),
\label{eq:hpe}
\end{equation}
and by defining $u=\left(\sqrt{\pi}a\right)\!/\!\left(\sqrt{2}W(R)\right)$:
\begin{equation}
h_{0}=\left[\mathrm{erf}(u)\right]^2, \quad W^2(R_{\mathrm{eq}}) = W^2(R)\frac{\sqrt{\pi}\mathrm{erf}(u)}{2u \exp(-u^2)},
\end{equation}
using the error function $\mathrm{erf}(u)=(2/\sqrt{\pi})\int_{0}^{u} \exp(-s^2)ds$.

\subsubsection{Atmospheric Turbulence}
\begin{figure}[t!]
    \centering
    \begin{subfigure}[t]{\linewidth}
        \centering
        \includegraphics[width=\linewidth]{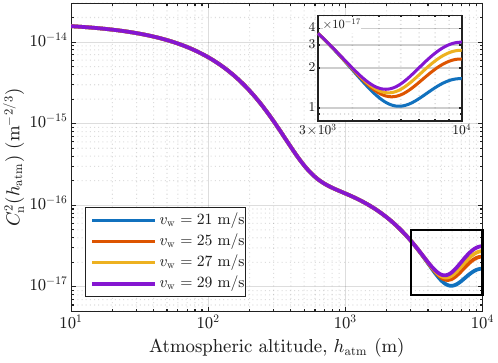}
        \caption{Altitude-dependent turbulence strength.\vspace{5pt}}
    \end{subfigure}
    \begin{subfigure}[t]{0.95\linewidth}
        \centering
        \includegraphics[width=\linewidth]{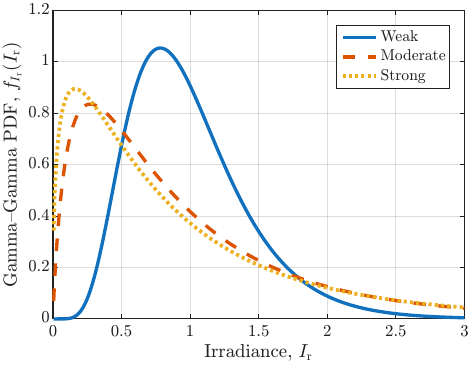}
        \caption{Gamma-Gamma irradiance PDF.}
    \end{subfigure}
    \caption{Turbulence-induced fading strength and statistics for weak, moderate, and strong turbulence conditions using $(\alpha_{\mathrm{GG}},\beta_{\mathrm{GG}})$ values of $(11.6,10.1)$, $(4,1.9)$ and $(4.2,1.4)$, respectively \cite{AlHabash2001GammaGamma}. As $\beta_{\mathrm{GG}} \rightarrow 1$, the irradiance statistics converge to a negative exponential distribution.\vspace{-10pt}}
    \label{fig:cn2_gamma}
\end{figure}

In \ac{LEO}-to-ground \ac{FSO} links, atmospheric turbulence causes random amplitude and phase distortions due to refractive-index fluctuations along the optical path \cite{andrews2001laser}. It can be classified into weak, moderate, or strong depending on the statistical strength of refractive-index variations. This is characterized using the structure parameter $C_{\mathrm{n}}^{2}(h_{\mathrm{atm}})$ as a function of atmospheric altitude $h_{\mathrm{atm}}$ based on the Hufnagel-Valley model \cite{andrews2001laser}:
\begin{equation}
\begin{aligned}
C_{\mathrm{n}}^{2}&(h_{\mathrm{atm}})
= 5.94\times10^{-3}
\!\left(\frac{v_{\mathrm{w}}}{27}\right)^{\!2}
(10^{-5}h_{\mathrm{atm}})^{10} \exp\!\left(\!-\frac{h_{\mathrm{atm}}}{1000}\!\right)\\
& +\,2.7\times10^{-16}\exp\!\left(\!-\frac{h_{\mathrm{atm}}}{1500}\!\right)
+\,A_{\mathrm{HV}}\exp\!\left(\!-\frac{h_{\mathrm{atm}}}{100}\!\right),
\end{aligned}
\label{eq:HV_mod}
\end{equation}
where $v_{\mathrm{w}}$ denotes the \ac{RMS} wind speed, and $A_{\mathrm{HV}}$ is the nominal near-ground turbulence constant. Irradiance fluctuations are modeled by the Gamma-Gamma distribution, which accurately describes the irradiance statistics for weak to strong turbulence conditions, with the \ac{PDF} \cite{AlHabash2001GammaGamma}:
\begin{equation}
\begin{aligned}
f_{I_{\mathrm{r}}}(I_{\mathrm{r}}) 
= {} & \frac{2(\alpha_{\mathrm{GG}}\beta_{\mathrm{GG}})^{(\alpha_{\mathrm{GG}}+\beta_{\mathrm{GG}})/2}}
        {\Gamma(\alpha_{\mathrm{GG}})\Gamma(\beta_{\mathrm{GG}})}
        \, I_{\mathrm{r}}^{(\alpha_{\mathrm{GG}}+\beta_{\mathrm{GG}})/2-1}  \\
& \times K_{\vartheta}\!\left(2\sqrt{\alpha_{\mathrm{GG}}\beta_{\mathrm{GG}} I_{\mathrm{r}}}\right),
\qquad I_{\mathrm{r}}>0 .
\end{aligned}
\label{eq:gamma_gamma_pdf}
\end{equation}
where $K_{\vartheta}$ denotes the modified Bessel function of the second kind and order $\vartheta$ for $\vartheta=\alpha_{\mathrm{GG}}-\beta_{\mathrm{GG}}$, and $\alpha_{\mathrm{GG}}$ and $\beta_{\mathrm{GG}}$ represent the effective numbers of large-scale and small-scale turbulent eddies, respectively. Under plane wave propagation, these parameters are related to the scintillation index $\sigma_{\mathrm{I}}^{2}$ (i.e., the normalized irradiance variance) via \cite{AlHabash2001GammaGamma}:
\begin{subequations}
\begin{align}
\alpha_{\mathrm{GG}} &=
\left[
\exp\!\left(
\frac{0.49\sigma_{\mathrm{I}}^{2}}
{\left(1+1.11\sigma_{\mathrm{I}}^{12/5}\right)^{\!7/6}}
\right) - 1
\right]^{-1},\\
\beta_{\mathrm{GG}} &=
\left[
\exp\!\left(
\frac{0.51\sigma_{\mathrm{I}}^{2}}
{\left(1+0.69\sigma_{\mathrm{I}}^{12/5}\right)^{\!5/6}}
\right) - 1
\right]^{-1}.
\end{align}
\end{subequations}
where \cite{AndrewsPhillips2005}:
\begin{equation}
\sigma_{\mathrm{I}}^{2}
\simeq 
\exp\!\left[
\frac{0.49\sigma_{\mathrm{R}}^{2}}
{\left(1+1.11\sigma_{\mathrm{R}}^{12/5}\right)^{\!7/6}}
+
\frac{0.51\sigma_{\mathrm{R}}^{2}}
{\left(1+0.69\sigma_{\mathrm{R}}^{12/5}\right)^{\!5/6}}
\right] - 1,
\label{eq:scint_index}
\end{equation}
where $\sigma_{\mathrm{R}}^{2}$ is the Rytov variance, representing the log-amplitude variance of the optical wave. Fig.~\ref{fig:cn2_gamma} illustrates the structure parameter $C_{\mathrm{n}}^{2}(h_{\mathrm{atm}})$ and the Gamma-Gamma irradiance \ac{PDF} under weak, moderate, and strong turbulence conditions. In \eqref{eq:fso_passband_complex}, the turbulence-induced phase $\phi_{\mathrm{turb}}(t)$ is modeled as a Wiener process to incorporate the cumulative nature of random wavefront distortion into the received optical field \cite{li2019phase}.

\subsection{Coherent Optical Receiver} \label{sec:cohRx}
The incident laser beam is collected by an \ac{FSO} telescope and coupled to a fiber using \ac{PAT}. In a coherent receiver, the received optical field $E_{\mathrm r}(t)$ is mixed with the \ac{LO} field $E_{\mathrm{LO}}(t)=\exp\!\left\{j\left[2\pi f_{\mathrm{LO}}t+\phi_{\mathrm{rx}}(t)\right]\right\}$ using a $90^\circ$ $2\times4$ hybrid comprising two 3-dB couplers and a $\pi/2$ phase shifter, as shown in Fig.~\ref{fig:system}. It generates four optical outputs with relative \ac{LO} phase shifts of $0$, $\pi/2$, $\pi$, and $3\pi/2$:
\begin{equation}
E_{m}(t)
=\frac{E_{\mathrm r}(t)
+E_{\mathrm{LO}}(t)e^{j\frac{\pi}{2}(m-1)}}{2},
\quad m\in\{1,2,3,4\}
\label{eq:A1}
\end{equation}
The opposite-phase outputs of the hybrid are fed into balanced \acp{PD}. After optical-to-electrical conversion, the resulting photocurrent differences in the in-phase and quadrature arms are obtained as $i_{\mathrm{I}}(t) = R_{\mathrm{PD}}\!\left(|E_{1}(t)|^{2}-|E_{3}(t)|^{2}\right)$ and $i_{\mathrm{Q}}(t) = R_{\mathrm{PD}}\!\left(|E_{2}(t)|^{2}-|E_{4}(t)|^{2}\right)$, respectively, where $R_{\mathrm{PD}}$ denotes the \ac{PD} responsivity. By expanding the squared magnitudes, it follows that:
\begin{subequations}
\begin{align}
i_{\mathrm{I}}(t)
&=\frac{R_{\mathrm{PD}}}{2}
\left(E_{\mathrm r}(t)E_{\mathrm LO}^{*}(t)
+E_{\mathrm r}^{*}(t)E_{\mathrm LO}(t)\right), \\
i_{\mathrm{Q}}(t)
&=\frac{R_{\mathrm{PD}}}{2}
\left(-jE_{\mathrm r}(t)E_{\mathrm LO}^{*}(t)
+jE_{\mathrm r}^{*}(t)E_{\mathrm LO}(t)\right).
\end{align}
\label{eq:A3}
\end{subequations}
By substituting \eqref{eq:fso_passband_complex} in \eqref{eq:A3}, using \eqref{eq:doppler_phase_shift}, and defining: 
\begin{equation}
\phi_{\mathrm{CFO}}(t)=2\pi\int_{0}^{t}\Delta f(\xi)\,\mathrm{d}\xi+\phi_{\mathrm{turb}}(t)+\phi_{\mathrm{PN}}(t),
\end{equation}
with $\Delta f(t)=f_{\mathrm{c}}-f_{\mathrm{LO}}+f_{\mathrm{D}}(t)+f_{\mathrm{l}}$ denoting the time-varying frequency offset, and $\phi_{\mathrm{PN}}(t)=\phi_{\mathrm{tx}}(t)-\phi_{\mathrm{rx}}(t)$, it can be shown that:
\begin{subequations}
\begin{align}
i_{\mathrm{I}}(t)
& \propto x_{\mathrm{I}}(t-\tau) \cos\phi_{\mathrm{CFO}}(t)-x_{\mathrm{Q}}(t-\tau)\sin\phi_{\mathrm{CFO}}(t), \\
i_{\mathrm{Q}}(t)
& \propto x_{\mathrm{I}}(t-\tau) \sin\phi_{\mathrm{CFO}}(t)+x_{\mathrm{Q}}(t-\tau)\cos\phi_{\mathrm{CFO}}(t),
\end{align}
\label{eq:hybIQ}
\end{subequations}
After matched filtering, the I and Q photocurrent signals are amplified by a \ac{TIA} and digitized by an \ac{ADC}. The Gram-Schmidt orthogonalization is then applied to remove any remaining \ac{IQ} imbalance \cite{coherbook}. This results in decorrelated discrete-time samples $r_\mathrm{I}[n]$ and $r_\mathrm{Q}[n]$, constituting the received complex baseband signal $r[n]=r_\mathrm{I}[n]+jr_\mathrm{Q}[n]$, which serves as input to carrier frequency and phase recovery algorithms.

\section{Frequency Offset Estimation Using the Hybrid Time-Frequency Method} \label{sec:method}
\subsection{4th-Power FFT-Based Coarse CFO Compensation}
The combined effects of the Doppler shift and atmospheric turbulence lead to a large frequency offset and slowly varying amplitude in the received optical field. To accurately retrieve the transmitted data, the frequency offset needs to be accurately estimated and counterbalanced using \ac{DSP}. To this end, first, the coarse frequency offset component is extracted based on the 4th-power \ac{FFT} spectrum to provide an initial estimate. By downconverting the received optical field in \eqref{eq:fso_passband_complex} to baseband followed by sampling, as discussed in Section~\ref{sec:cohRx}, and timing synchronization, the discrete-time complex baseband received signal is obtained as:
\begin{equation}
\begin{aligned}
r[n] = {} & G[n]\sqrt{I_\mathrm{r}[n]}\,x[n]
\exp\!\Big\{ j\big(\Phi_{\mathrm{D}}[n]
+ \phi_{\mathrm{turb}}[n]+\phi_{\mathrm{PN}}[n]\big) \Big\} \\
& + z[n],
\end{aligned}
\label{eq:bb_turb_cfo_full}
\end{equation}
where $G[n]=\sqrt{\eta_{\mathrm{tx}}\eta_{\mathrm{rx}}P_\mathrm{t}L_{\mathrm{geo}}(R)L_{\mathrm{atm}}(R)h_{\mathrm{pe}}[n]}$ represents deterministic path loss and quasi-static pointing error effects, which are approximately constant within a symbol interval so that $G[n]\approx G$. During a full \ac{LEO} pass, the Doppler-induced \ac{CFO} phase term $\Phi_{\mathrm{D}}[n]$ can be described by a cumulative time-varying process \cite{shoji2012pilot}:
\begin{equation}
\Phi_{\mathrm{D}}[n]
=\Phi_{\mathrm{D}}[n-1]+2\pi \Delta f[n]T_{\mathrm{s}}
=2\pi\sum_{k=0}^{n}\Delta f[k]T_{\mathrm{s}},
\label{eq:cumulative_cfo_phase}
\end{equation}
where $\Delta f[n]$ is the sampled version of $\Delta f(t)$. Using a short pilot acquisition window of $\mathcal{W}_{\mathrm{p}}=\{n_0,\ldots,n_0+N_{\mathrm{p}}-1\}$ with length $N_{\mathrm{p}}$, the Doppler frequency is locally approximated as a slowly varying process \cite{shoji2012pilot}. Therefore:
\begin{equation}
\Phi_{\mathrm{D}}[n_0+m]
=\Phi_{\mathrm{D}}[n_0]+2\pi\Delta f[n_0]mT_{\mathrm{s}}
+\delta_{\mathrm{D}}[m],
\label{eq:local_cfo_phase}
\end{equation}
for $m=0,\ldots,N_{\mathrm{p}}-1$, where:
\begin{equation}
\delta_{\mathrm{D}}[m]=2\pi\sum_{q=1}^{m}\big(\Delta f[n_0+q]-\Delta f[n_0]\big)T_{\mathrm{s}},
\label{eq:Doppler_var}
\end{equation}
is the intra-window Doppler variation. For a sufficiently small value of $|\delta_{\mathrm{D}}[m]|$, the \ac{CFO} can be locally approximated by its instantaneous value $\Delta f[n_0]$ for coarse acquisition. To this end, a useful design condition follows from the first-order Doppler-rate approximation: $\Delta f[n_0+q]-\Delta f[n_0]\approx qT_{\mathrm{s}}\dot{f}_{\mathrm{D}}[n_0]$, where $\dot{f}_{\mathrm{D}}[n_0]=df_{\mathrm{D}}(t)/dt\big|_{t=n_0T_s}$, which results in:
\begin{equation}
\max_{0\leq m < N_{\mathrm{p}}}|\delta_{\mathrm{D}}[m]|
\lesssim
\pi |\dot{f}_{\mathrm{D}}[n_0]|T_{\mathrm{s}}^{2}N_{\mathrm{p}}(N_{\mathrm{p}}-1)
\ll 1 .
\label{eq:local_doppler_condition}
\end{equation}

In \ac{QPSK}, by raising \eqref{eq:bb_turb_cfo_full} to the power of $4$, the data symbol term becomes $x^4[n]$, which is a constant-phase factor and can be discarded, leading to:
\begin{equation}\label{eq:fourth_power_turb}
\begin{aligned}
r^{4}[n] \approx {} & G^{4} I_\mathrm{r}^{2}[n]\,
\exp\!\left\{j4\Phi_{\mathrm{D}}[n]
+j\,4(\phi_{\mathrm{turb}}[n]+\phi_{\mathrm{PN}}[n])\right\} \\
& + \tilde{z}[n].
\end{aligned}
\end{equation}
where $\tilde{z}[n]$ is the effective noise arsing from the 4th-power operation. Under the approximation in \eqref{eq:local_cfo_phase}, the spectrum of $r^{4}[n]$ over the acquisition window $\mathcal{W}_{\mathrm{p}}$ contains a dominant tone at $4\Delta f[n_0]$ within the \ac{FFT} observation range $[-F_{\mathrm{s}}/2,F_{\mathrm{s}}/2)$ with a small spectral broadening caused by the intra-window Doppler variation in \eqref{eq:Doppler_var}. The alias-free, unambiguous coarse \ac{CFO} estimation range is obtained from the Nyquist condition $4|\Delta f[n_0]| < F_{\mathrm{s}}/2$, which means $|\Delta f[n_0]| < F_{\mathrm{s}}/8$. If the true \ac{CFO} falls outside this range, a straightforward solution is to increase the sampling frequency $F_{\mathrm{s}}$. Alternatively, the 4th-power tone $4|\Delta f[n_0]|$ can be folded into the \ac{FFT} range using a modulo operation: $\left(\left(4\Delta f[n_0]+F_{\mathrm{s}}/2\right)\bmod F_{\mathrm{s}}\right)-F_{\mathrm{s}}/2$.

For higher-order modulations such as \ac{16QAM}, the 4th-power operation produces a data-dependent pedestal around the desired tone. The spectral peak is detectable as long as the intra-window Doppler shift and turbulence-induced irradiance and phase are slowly varying with respect to the \ac{FFT} length. Once the dominant bin near $4\Delta f[n_0]$ is found, the off-bin location of the spectral peak is refined using parabolic interpolation. To this end, the 4th-power spectrum is locally approximated by a second-order polynomial in the neighborhood of the dominant bin. Let $\hat{S}[k] = S(f_k)$ and $S(f)$ be the magnitude spectrum of $r^{4}[n]$ and its continuous counterpart, respectively, for $f_k = kF_\mathrm{s}/N_{\mathrm{FFT}}$, where $N_{\mathrm{FFT}}$ is the \ac{FFT} size. Assuming that $S(f)$ is locally twice differentiable, the Taylor expansion around the discrete maximizer $f_{k_0}$ is:
\begin{equation}
    S(f) \approx S(f_{k_0}) 
  + S'(f_{k_0})(f-f_{k_0})
  + \tfrac{1}{2}S''(f_{k_0})(f-f_{k_0})^{2}.
  \label{eq:Sf_approx}
\end{equation}
Setting the derivative of \eqref{eq:Sf_approx} to zero gives rise to:
\begin{equation}
    f_{\mathrm{interp}} = f_{k_0} - \frac{S'(f_{k_0})}{S''(f_{k_0})},
    \label{eq:finterp}
\end{equation}
where:
\begin{subequations}\label{eq:interp_derivatives}
\begin{align}
S'(f_{k_0}) &\approx 
  \frac{\hat{S}[k_0+1]-\hat{S}[k_0-1]}{2f_\mathrm{s}}, \\[4pt]
S''(f_{k_0}) &\approx
  \frac{\hat{S}[k_0+1]-2\hat{S}[k_0]+\hat{S}[k_0-1]}{f_\mathrm{s}^{2}},
\end{align}
\end{subequations}
where $f_\mathrm{s} = F_\mathrm{s} / N_{\mathrm{FFT}}$. The interpolated peak frequency in \eqref{eq:finterp} provides a higher-resolution estimate of the detected tone over the alias-free interval $|\Delta f[n_0]| < F_{\mathrm{s}}/8$. The 4th-power coarse \ac{CFO} estimation procedure is summarized in Algorithm~\ref{alg:coarseCFO}. The function $\mathrm{fftshift}$ centers data along the \ac{FFT} frequency axis.

\begin{algorithm}[t!]
\caption{4th-Power \ac{FFT}-Based Coarse \ac{CFO} Estimation}
\label{alg:coarseCFO}
\begin{algorithmic}[1]
\renewcommand{\algorithmicrequire}{\textbf{Input}:}
\renewcommand{\algorithmicensure}{\textbf{Output}:}
\Require Received signal $r[n]$, pilot length $N_\mathrm{p}$, symbol period $T_{\mathrm{s}}$, and sampling frequency $F_{\mathrm{s}}$
\Ensure Coarse \ac{CFO} estimate $\widehat{\Delta f}_{\mathrm{coarse}}$ and coarse-corrected samples $r_{\mathrm{coarse}}[n]$
\State Select the first $N_\mathrm{p}$ samples: $r_{\mathrm{p}} \gets r[0:N_\mathrm{p}-1]$
\State Perform the 4th-power operation and find the \ac{FFT} size: \[y_4[n] \gets r_{\mathrm{p}}^4[n], \quad N_{\mathrm{FFT}} \gets 2^{\lceil \log_2 N_\mathrm{p} \rceil}\]
\State Compute the \ac{FFT} frequency spacing: $f_\mathrm{s}\gets \dfrac{F_{\mathrm{s}}}{N_{\mathrm{FFT}}}$
\State Compute the \ac{FFT} and the magnitude spectrum: \[Y_4 \gets \mathrm{FFT}(y_4,N_{\mathrm{FFT}}), \quad \hat{S} \gets |\mathrm{fftshift}(Y_4)|\]
\State Locate the spectral peak: $k_0 \gets \arg\underset{k}{\max}\ \hat{S}[k]$
\State Apply the parabolic interpolation:
\If{$1<k_0<N_{\mathrm{FFT}}$}
\State $\displaystyle \delta_k \gets \frac{1}{2}\,
\frac{\hat{S}[k_0-1]-\hat{S}[k_0+1]}
{\hat{S}[k_0-1]-2\hat{S}[k_0]+\hat{S}[k_0+1]}$
\Else
\State $\delta_k \gets 0$
\EndIf
\State Find the interpolated frequency index: $k_{\mathrm{interp}} \gets k_0 + \delta_k$
\State Convert the \ac{FFT} index to frequency after $\mathrm{fftshift}$: \[f_4 \gets \left(k_{\mathrm{interp}} - 1 - \frac{N_{\mathrm{FFT}}}{2}\right) f_\mathrm{s}\]
\State Compute the coarse \ac{CFO} estimate: $\widehat{\Delta f}_{\mathrm{coarse}} \gets \dfrac{f_4}{4}$
\State Perform the coarse \ac{CFO} correction: \[r_{\mathrm{coarse}}[n] \gets r[n]\exp\!\left(-j2\pi \widehat{\Delta f}_{\mathrm{coarse}} T_{\mathrm{s}} n\right)\]
\end{algorithmic}
\end{algorithm}

\begin{algorithm}[t!]
\caption{DD-FLL Stability and Capture Feasibility Check}
\label{alg:stability_check}
\begin{algorithmic}[1]
\renewcommand{\algorithmicrequire}{\textbf{Input}:}
\renewcommand{\algorithmicensure}{\textbf{Output}:}
\Require Coarse-corrected samples $r_{\mathrm{coarse}}[n]$, handover length $N_{\mathrm{ho}}$, pilot or decision symbols $\tilde{x}[n]$, symbol period $T_{\mathrm{s}}$, symbol rate $R_{\mathrm{s}}=1/T_{\mathrm{s}}$, residual offset design parameter $f_{\max}$, handover margin $\gamma_{\mathrm{ho}}$, and lock margin $\epsilon_{\mathrm{lock}}$
\Ensure Feasibility flag $\mathsf{Lock}$ and residual frequency initialization $\omega[0]$
\State Compute the tracking limit: \[f_{\mathrm{tr}} \gets 2f_{\max}, \quad \omega_{\max} \gets \gamma_{\mathrm{ho}}\frac{2\pi f_{\mathrm{tr}}}{R_{\mathrm{s}}}\]
\State Estimate the residual angular frequency offset after coarse correction:
\[\hat{\omega}_{\mathrm{res},0}\gets\angle\!\left\{\sum_{n=2}^{N_{\mathrm{ho}}}\frac{r_{\mathrm{coarse}}[n]\,r_{\mathrm{coarse}}^{*}[n-1]}{\tilde{x}[n]\tilde{x}^{*}[n-1]}\right\}\]
\State Define the admissible region: $\mathcal{A}_{\mathrm{cap}}: |\hat{\omega}_{\mathrm{res},0}| \le \epsilon_{\mathrm{lock}}\,\omega_{\max}$
\If{$\hat{\omega}_{\mathrm{res},0} \in \mathcal{A}_{\mathrm{cap}}$}: Stable handover is feasible
    \State $\mathsf{Lock}\gets 1, \quad \omega[0]\gets \hat{\omega}_{\mathrm{res},0}$
\Else: Out of capture (i.e., risk of divergence)
    \State $\mathsf{Lock}\gets 0, \quad \omega[0]\gets \mathrm{clip}\big(\hat{\omega}_{\mathrm{res},0},-\omega_{\max},\omega_{\max}\big)$
    \State Re-run coarse acquisition using a larger $N_\mathrm{p}$ or refined peak search based on Algorithm~\ref{alg:coarseCFO}
\EndIf
\State The equivalent residual \ac{CFO} bound: $|\Delta f_{\mathrm{res}}| \le \dfrac{\omega_{\max}}{2\pi T_{\mathrm{s}}}$
\end{algorithmic}
\end{algorithm}

\subsection{FLL-Based Residual CFO Compensation}
Upon completing coarse acquisition and correction, the \ac{DD}-\ac{FLL} feedback tracks the residual \ac{CFO}. Unlike the 4th-power coarse \ac{CFO} acquisition, the \ac{DD}-\ac{FLL} has a finite capture range for residual \ac{CFO} tracking. Therefore, after coarse correction, the \ac{CFO} capture feasibility needs to be checked to ensure that the residual \ac{CFO} lies within the reliable convergence range of the \ac{DD}-\ac{FLL}, similar to closed-loop tracking procedures in \ac{PLL} or \ac{AFC} acquisition \cite{messerschmitt1979frequency,natali1984afc}. Let $r_{\mathrm{coarse}}[n]$ be the output of Algorithm~\ref{alg:coarseCFO}. For a handover window of $N_{\mathrm{ho}}$ symbols, the phase increment due to the residual coarse \ac{CFO} is \cite{freq1}:
\begin{equation}
\hat{\omega}_{\mathrm{res},0}
=\angle\!\left\{
\sum_{n=2}^{N_{\mathrm{ho}}}
\frac{r_{\mathrm{coarse}}[n]\,r_{\mathrm{coarse}}^{*}[n-1]}
{\tilde{x}[n]\tilde{x}^{*}[n-1]}
\right\},
\label{eq:residual_handover}
\end{equation}
where $\angle$ represents the polar angle of a complex variable, and $\tilde{x}[n]$ is either the known pilot symbol or the hard decision after coarse correction. Checking the \ac{DD}-\ac{FLL} feasibility is carried out by Algorithm~\ref{alg:stability_check}. The tracking loop imposes the admissible residual \ac{CFO} bound: 
\begin{equation}
|\Delta f_{\mathrm{res}}|\le \frac{\omega_{\max}}{2\pi T_{\mathrm{s}}}, \quad \omega_{\max}=\gamma_{\mathrm{ho}}\frac{2\pi f_{\mathrm{tr}}}{R_{\mathrm{s}}}, \quad f_{\mathrm{tr}}=2f_{\max}
\label{eq:FLL_handover}
\end{equation}
where $f_{\max}$ sets the expected post-coarse correction frequency, and $\gamma_{\mathrm{ho}}$ is handover margin for discriminator noise and interpolation error. The activation condition in \eqref{eq:FLL_handover} is conservative, since Algorithm~\ref{alg:stability_check} requires $|\hat{\omega}_{\mathrm{res},0}|\le \epsilon_{\mathrm{lock}}\omega_{\max}$ for $\epsilon_{\mathrm{lock}}<1$. In practice, $f_{\max}$ is chosen to be above the \ac{RMS} coarse estimation error and below the symbol rate ambiguity limit, while ensuring $\epsilon_{\mathrm{lock}}\omega_{\max}<\pi$ to avoid phase increment wrapping. Algorithm~\ref{alg:stability_check} passes the clipped residual \ac{CFO} value to the \ac{DD}-\ac{FLL} for per-symbol frequency offset tracking. In Algorithm~\ref{alg:stability_check}, the function $\mathrm{clip}(z,z_{\min},z_{\max})=\min(\max(z,z_{\min}),z_{\max})$, which limits the initial loop frequency to $[z_{\min},z_{\max}]$.

The \ac{DD}-\ac{FLL} feedback tracks the residual \ac{CFO} based on the initial value $\omega[0]$ provided by Algorithm~\ref{alg:stability_check}. First, each coarse-corrected sample is derotated by the current phase estimate $\hat{\phi}_{\mathrm{FLL}}[n-1]$ to obtain $y[n]=r_{\mathrm{coarse}}[n]e^{-j\hat{\phi}_{\mathrm{FLL}}[n-1]}$. Then, a hard decision is applied to acquire the estimated symbol $\hat{x}[n]$. To estimate and cancel inter-symbol phase increment due to the residual \ac{CFO}, a \ac{DD} frequency discriminator is used \cite{freq1}:
\begin{equation}
    e_\mathrm{f}[n]
= \angle \left\{
\frac{y[n]\,y^{*}[n-1]}{\hat{x}[n]\hat{x}^{*}[n-1]}
\right\}.
\label{eq:discriminator}
\end{equation}
A first-order \ac{IIR} \ac{LPF} is used to suppress the decision noise before the second-order \ac{PI} loop updates the per-symbol angular frequency $\hat{\omega}[n]$ and phase $\hat{\phi}_{\mathrm{FLL}}[n]$. The described procedure is summarized in Algorithm~\ref{alg:fll}. The \ac{DD}-\ac{FLL} output $y[n]$ is passed on to carrier phase recovery and detection, while $\widehat{\Delta f}_{\mathrm{res}}[n]$ and $\widehat{\Delta f}_{\mathrm{tot}}[n]$ are used to evaluate the tracking performance. In Algorithm~\ref{alg:fll}, $\mathrm{dec}$ denotes the nearest-neighbor decoding for \ac{QPSK} or \ac{16QAM} constellations.

\begin{algorithm}[t!]
\caption{DD-\ac{FLL} Tracking Loop (2nd Order \ac{PI} Structure)}
\label{alg:fll}
\begin{algorithmic}[1]
\renewcommand{\algorithmicrequire}{\textbf{Input}:}
\renewcommand{\algorithmicensure}{\textbf{Output}:}
\Require Coarse estimate $\widehat{\Delta f}_{\mathrm{coarse}}$ and coarse-corrected samples $r_{\mathrm{coarse}}[n]$ from Algorithm~\ref{alg:coarseCFO}, residual frequency initialization $\omega[0]$ from Algorithm~\ref{alg:stability_check}, handover margin $\gamma_{\mathrm{ho}}$, symbol period $T_{\mathrm{s}}$, and symbol rate $R_{\mathrm{s}}=1/T_{\mathrm{s}}$
\Ensure Corrected samples $y[n]$, residual \ac{CFO} estimate $\widehat{\Delta f}_{\mathrm{res}}[n]$, and total \ac{CFO} estimate $\widehat{\Delta f}_{\mathrm{tot}}[n]$
\renewcommand{\algorithmicrequire}{\textbf{Parameters}:}
\Require Proportional gain $K_\mathrm{p}$, integral gain $K_\mathrm{i}$, low-pass factor $\alpha_{\mathrm{LP}}$, and residual offset design parameter $f_{\max}$
\State Initialize: \[\hat{\phi}_{\mathrm{FLL}}[0] \gets 0, \, \hat{\omega}[0] \gets \omega[0], \, e_{\mathrm{LP}}[0]\gets0, \, y[1]\gets r_{\mathrm{coarse}}[1]\]
\State Apply hard decision on the first symbol: $\hat{x}[1] = \mathrm{dec}(y[1])$
\State Compute the residual \ac{CFO} limit: $\omega_{\max} \gets \gamma_{\mathrm{ho}}\dfrac{4\pi f_{\max}}{R_{\mathrm{s}}}$
\For{$n=2$ to $N_{\mathrm{s}}$}
    \State Phase rotation: $y[n] \gets r_{\mathrm{coarse}}[n]\,e^{-j\hat{\phi}_{\mathrm{FLL}}[n-1]}$
    \State Symbol decision: $\hat{x}[n] = \mathrm{dec}(y[n])$
    \State Frequency discriminator:
    \[e_\mathrm{f}[n] \gets \angle \left\{\frac{y[n]\,y^{*}[n-1]}{\hat{x}[n]\hat{x}^{*}[n-1]}\right\}\]
    \State LPF: $e_{\mathrm{LP}}[n] \gets (1-\alpha_{\mathrm{LP}})e_{\mathrm{LP}}[n-1] + \alpha_{\mathrm{LP}} e_\mathrm{f}[n]$
    \State Angular frequency update: $\hat{\omega}[n] \gets \hat{\omega}[n-1] + K_\mathrm{i} e_{\mathrm{LP}}[n]$
    \State Limiter: $\hat{\omega}[n] \gets \mathrm{clip}\big(\hat{\omega}[n],-\omega_{\max},\omega_{\max}\big)$
    \State Phase update: \[\hat{\phi}_{\mathrm{FLL}}[n] \gets \hat{\phi}_{\mathrm{FLL}}[n-1] + \hat{\omega}[n] + K_\mathrm{p} e_{\mathrm{LP}}[n]\]
    \State Residual CFO estimate: $\widehat{\Delta f}_{\mathrm{res}}[n] \gets \dfrac{\hat{\omega}[n]}{2\pi T_{\mathrm{s}}}$
    \State Total CFO estimate: $\widehat{\Delta f}_{\mathrm{tot}}[n] \gets \widehat{\Delta f}_{\mathrm{coarse}}+\widehat{\Delta f}_{\mathrm{res}}[n]$
\EndFor
\State Return $\{y[n]\}_{n=1}^{N_{\mathrm{s}}}$, $\{\widehat{\Delta f}_{\mathrm{res}}[n]\}_{n=1}^{N_{\mathrm{s}}}$, and $\{\widehat{\Delta f}_{\mathrm{tot}}[n]\}_{n=1}^{N_{\mathrm{s}}}$
\end{algorithmic}
\end{algorithm}

\subsection{Computational Complexity}
Computational complexity is an important design metric for evaluating the feasibility of \ac{DSP} algorithms. For the 4th-power \ac{FFT}-based coarse estimator, let $N_\mathrm{p}$ denote the number of pilot symbols. From Algorithm~\ref{alg:coarseCFO}, the \ac{FFT} size is chosen as $N_{\mathrm{FFT}}=2^{\lceil\log_2 N_\mathrm{p}\rceil}$, where $\lceil s\rceil$ denotes the smallest integer satisfying $\lceil s\rceil\geq s$. Therefore, $N_{\mathrm{FFT}}\approx N_\mathrm{p}$ up to a zero-padding factor. The 4th-power preprocessing involves two complex multiplications per pilot symbol, equivalent to $\approx 8N_\mathrm{p}$ real multiplications. The radix-2 Cooley-Tukey \ac{FFT} costs $5N_{\mathrm{FFT}}\log_2 N_{\mathrm{FFT}}$ real multiplications, where the factor~$5$ accounts for the butterfly arithmetic cost $2N_{\mathrm{FFT}}\log_2 N_{\mathrm{FFT}}$ with an overhead of $\approx3N_{\mathrm{FFT}}\log_2 N_{\mathrm{FFT}}$ for twiddle factors, memory access and control \cite{garrido2018hardware}. Computing the magnitude spectrum and the peak frequency need $2N_{\mathrm{FFT}}$ real multiplications. Moreover, coarse \ac{CFO} correction uses one complex multiplication, or $\approx 4N_{\mathrm{d}}$ real multiplications, where $N_{\mathrm{d}}$ is the number of data symbols. Hence, the total computational cost of coarse \ac{CFO} estimation and correction in Algorithm~\ref{alg:coarseCFO} is:
\begin{equation}
\label{eq:complex_coarse}
M_{\mathrm{coarse}} = \mathcal{O}\big(5N_\mathrm{p} \log N_\mathrm{p} + 10N_\mathrm{p} + 4N_{\mathrm{d}}\big).
\end{equation}

In Algorithm~\ref{alg:fll}, the \ac{DD}-\ac{FLL} processes every data symbol, thereby increasing complexity linearly with $N_{\mathrm{d}}$. For each symbol, the phase rotation requires one complex or four real
multiplications. The frequency discriminator executes three complex multiplications and a unit-modulus division, resulting in about $12$ real multiplications. The \ac{IIR} \ac{LPF} adds two real multiplications, and the \ac{PI} controller contributes another two real multiplications for angular frequency integration and proportional phase update. Thus, the total cost per symbol is $20$ real multiplications. Consequently, the total computational cost of the proposed \ac{CFO} tracking algorithm is:
\begin{equation}
M_{\mathrm{total}} = \mathcal{O} \big(5N_\mathrm{p} \log N_\mathrm{p} + 10N_\mathrm{p} + 24N_{\mathrm{d}}\big).
\end{equation}

\subsection{Phase Noise Compensation Using Carrier Phase Recovery}
\begin{figure}[t!]
\centering
\includegraphics[width=1\linewidth,trim=176bp 129bp 168bp 90bp,clip]{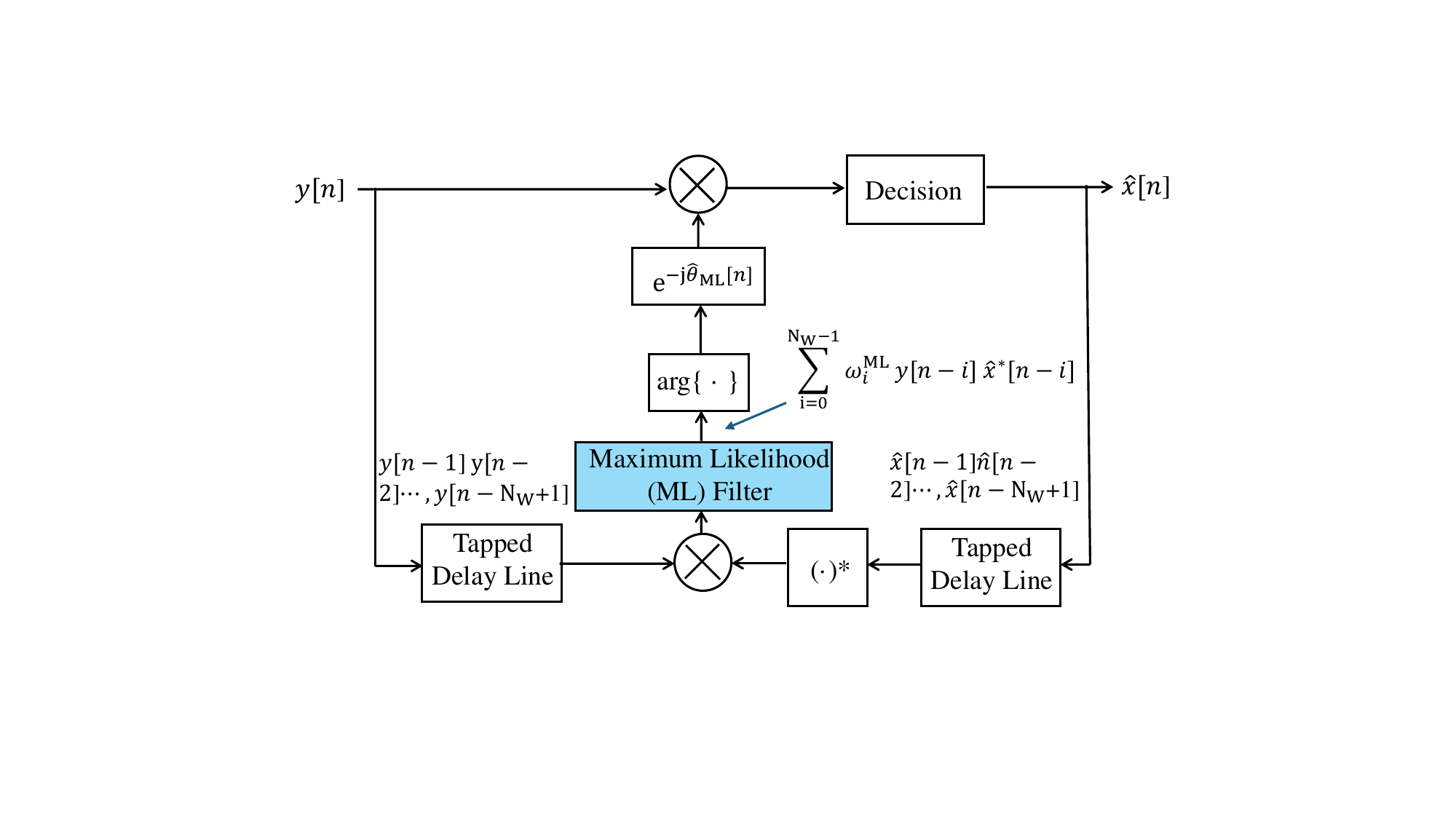}
\caption{DD-CPR with tapped-delay \ac{ML} filter.\vspace{-10pt}}
\label{fig:ML_PR}
\end{figure}

A \ac{DD}-\ac{CPR} algorithm is used to estimate and compensate for the phase noise $\phi_{\mathrm{PN}}[n]$ using an \ac{ML} filter~\cite{coherbook}. Fig.~\ref{fig:ML_PR} shows the block diagram of the \ac{ML}-based phase tracker. A sequence of $N_{\mathrm{s}}$ received symbols $\mathbf{y} = [y[n],y[n-1],\dots,y[n-N_{\mathrm{s}}+1]]^T$ is processed, and the resulting hard decisions $\hat{\mathbf{x}}=[\hat x[n],\hat x[n-1],\dots,\hat x[n-N_{\mathrm{s}}+1]]^T$ are fed back for phase tracking. The instantaneous phase estimate is computed as~\cite{coherbook}:
\begin{equation}
\hat{\theta}_{\mathrm{ML}}[n]
=
\angle \left\{ \sum_{i=0}^{N_{\mathrm{s}}-1}
w_{i}^{\mathrm{ML}}\,y[n-i]\hat x^*[n-i] \right\}.
\end{equation}
The \ac{ML} tap coefficient vector $\mathbf{w}_{\mathrm{ML}}=[w_{i}^{\mathrm{ML}}]_{i=0,1,\dots,N_{\mathrm{s}}-1}$ is obtained as $\mathbf{w}_{\mathrm{ML}} = \frac{\mathbf{C}^{-1}\mathbf{1}}{\mathbf{1}^{\mathsf T}\mathbf{C}^{-1}\mathbf{1}}$, where $\mathbf{1}$ denotes the all-ones vector, and $\mathbf{C}$ is the covariance matrix of the vector of the decision-aided products $y[n]\hat x^{*}[n]$, which can be expressed as:
\begin{equation}
\mathbf{C} = \zeta_{\mathrm{PN}}\,E_\mathrm{s}^2\,\mathbf{K}_{\mathrm{PN}} + \zeta_{\mathrm{n}}\,E_\mathrm{s}\,\mathbf{I},
\label{eq:cov_factored}
\end{equation}
where $E_\mathrm{s}$ is the average symbol energy, $\zeta_{\mathrm{PN}}$ and $\zeta_{\mathrm{n}}$ are the phase noise and additive noise scaling factors, respectively, $\mathbf{K}_{\mathrm{PN}}$ is the covariance of the accumulated phase noise within a tapped delay window, and $\mathbf{I}$ is the identity matrix. According to a Wiener process model for the phase noise, the phase sample for a given tap equals the cumulative sum of independent phase increments up to that tap. This creates a lower-triangular covariance matrix with entries $[\mathbf{K}_{\mathrm{PN}}]_{i,j}=\min(i,j)+1$ for $i,j=0,\dots,N_{\mathrm{s}}\!-\!1$. Following the \ac{DD}-\ac{CPR}, by quantizing the \ac{IQ} components of the received phase-compensated symbol, a hard decision is made, and the phase tracking buffer is updated with the normalized product of $y[n]$ and the decided symbol.

\section{Pairwise Error Probability Analysis\\ Under the Residual CFO}\label{DERIVATION OF PEP UNDER RESIDUAL CFO} \label{sec:pep}
The residual \ac{CFO} affects the received signal constellation and can lead to symbol decision errors. \ac{PEP} is an effective performance metric to evaluate the impact of such constellation distortions. The \ac{PEP} analysis is performed by finding the conditional \ac{PEP} given the post-acquisition residual phase, then averaging it over the phase trajectory associated with the residual \ac{CFO}. To elaborate, recall the cumulative time-varying phase model from \eqref{eq:cumulative_cfo_phase}. For a residual \ac{CFO} trajectory $\Delta f_{\mathrm{res}}[n]$ after compensation, the corresponding residual phase sequence evolves as \cite{kurz2013recursive}:
\begin{equation}
    \phi_{\mathrm{res}}[n] = \Theta_{2\pi}\big(\phi_{\mathrm{res}}[n-1]+2\pi \Delta f_{\mathrm{res}}[n]T_{\mathrm{s}}+\epsilon_{\mathrm{CPR}}[n]\big),
\end{equation}
where $\Theta_{2\pi}(\phi)$ is a phase wrapping function that bounds $\phi$ to a $2\pi$ interval, and $\epsilon_{\mathrm{CPR}}[n]$ is the \ac{CPR} error. For a given \ac{CFO} trajectory using an observation window $\mathcal{W}_{\mathrm{p}}$ of length $N_{\mathrm{p}}$, the empirical \ac{PDF} of the residual phase is given by \cite{kurz2013recursive}:
\begin{equation}
f_{\phi}(\phi)
 = \frac{1}{N_{\mathrm{p}}}
   \sum_{n\in\mathcal{W}_{\mathrm{p}}}
   \delta_{2\pi}(\phi-\phi_{\mathrm{res}}[n]),
\label{eq:empirical_phase_pdf}
\end{equation}
where $\delta_{2\pi}$ denotes an impulse train with a period of $2\pi$. A simplified notation of $\phi$ is used for the random residual phase. 

Recall the complex baseband received signal model from \eqref{eq:bb_turb_cfo_full}. The received signal sample is:
\begin{equation}
 r = h X e^{j\phi}+z,
 \label{eq:rxPEP}
\end{equation}
where $h=G h_{\mathrm{turb}}$ and $G$ is a constant attenuation factor, and $z\sim\mathcal{CN}(0,N_{0})$ is an \ac{AWGN} with variance $N_0=E_{\mathrm{s}}/(\gamma_{\mathrm{b}}\log_2M)$, where $\gamma_{\mathrm{b}}=E_{\mathrm{b}}/N_0$ denotes the \ac{SNR} per bit, and $M$ is the constellation size. Consider a pairwise error event in which $X$ is transmitted but $\hat{X}$ is detected. The receiver uses an \ac{ML} minimum Euclidean distance criterion for symbol demapping. For a given symbol $\Omega$ chosen from the alphabet $\mathcal{X}$, the \ac{ML} decision metric is \cite{Proakis2008DigitalCommunications}:
\begin{equation}
    \Lambda(\Omega) = 2\Re\big\{r^{*}h\Omega\big\} - |h|^2|\Omega|^2,
    \label{eq:ML_metric}
\end{equation}
where $^{*}$ denotes the complex conjugate operator. Let $\mathbb{P}$ denote the probability of an event. The conditional \ac{PEP} given $\phi$ and $|h|$ is:
\begin{equation}
\begin{aligned}
    \mathbb{P}\big\{X\rightarrow \hat{X} \mid \phi,|h|\big\} 
    &= \mathbb{P}\big\{ \Lambda(\hat{X}) > \Lambda(X) \mid \phi,|h| \big\}, \\
    &= \mathbb{P}\big\{ \Lambda(\hat{X})-\Lambda(X) > 0 \mid \phi,|h| \big\}.
\end{aligned}
\label{eq:conditional_PEP}
\end{equation}
Based on \eqref{eq:rxPEP} and \eqref{eq:ML_metric}:
\begin{equation}
\begin{aligned}
    \Lambda(\hat{X})-\Lambda(X) &= 2\Re\big\{r^{*} h(\hat{X}-X)\big\} - |h|^2\big(|\hat{X}|^2-|X|^2\big), \\
    &= \Psi\big(\phi,|h|\big) + \nu,
\end{aligned}
\label{eq:ML_metricdiff}
\end{equation}
where:
\begin{equation}
    \Psi\big(\phi,|h|\big) = 2\Re\{X^{*}e^{-j\phi}|h|^{2}(\hat{X}-X)\} - |h|^2\big(|\hat{X}|^2-|X|^2\big), 
    \label{eq:Psi_PEP}
\end{equation}
and:
\begin{equation}
    \nu = 2\Re\{hz^{*}(\hat{X}-X)\}, 
    \label{eq:nu_PEP}
\end{equation}
is the noise term. Note that: 
\begin{equation}
    hz^{*}(\hat{X}-X)\sim\mathcal{CN}\big(0,|h|^{2}N_0|\hat{X}-X|^2\big).
\end{equation}
Therefore, the noise term $\nu \sim \mathcal{N}\!\left(0,\sigma_{\nu}^{2}\right)$, where:
\begin{equation}
    \sigma_{\nu}^2 = \mathbb{E}\big\{\nu^{2}\big\}= 2 |h|^{2} N_0 \big|\hat{X}-X\big|^{2},
    \label{eq:varnu_PEP}
\end{equation}
where $\mathbb{E}$ denotes the expectation operator. By combining \eqref{eq:conditional_PEP} with \eqref{eq:ML_metricdiff}--\eqref{eq:nu_PEP}, the conditional \ac{PEP} is then derived as:
\begin{equation}
\begin{aligned}
    \mathbb{P}\big\{X\rightarrow\hat{X} \mid \phi,|h|\big\} &= \mathbb{P}\big\{ \Psi(\phi,|h|) + \nu > 0 \mid \phi,|h| \big\}, \\
    &= \mathbb{P}\big\{\nu > -\Psi(\phi,|h|) \mid \phi,|h| \big\}, \\
    &= Q\!\left(-\frac{\Psi\big(\phi,|h|\big)}{\sigma_{\nu}} \right),
\end{aligned}
    \label{eq:simplified_PEP}
\end{equation}
where $Q(u)=\frac{1}{\sqrt{2\pi}}\int_{u}^{\infty}\exp\left(-s^2/2\right)ds$ is the $Q$-function. Substituting \eqref{eq:Psi_PEP} and \eqref{eq:varnu_PEP} in \eqref{eq:simplified_PEP} gives rise to:
\begin{equation}
\mathbb{P}\big\{X\rightarrow\hat{X}\mid \phi,|h|\big\} = Q\!\left(\Gamma_{\phi}(X,\hat{X})\right),
\label{eq:generalized_PEP}
\end{equation}
where:
\begin{equation}
\Gamma_{\phi}(X,\hat{X})
= |h|\, \frac{|\hat{X}|^2-|X|^2 - 2\Re\big\{X^{*}e^{-j\phi}(\hat{X}-X)\big\}}{\sqrt{2N_0}\big|\hat{X}-X\big|}.
\label{eq:gamma_phi_generalized}
\end{equation}
The conditional \ac{PEP} presented in \eqref{eq:generalized_PEP} along with \eqref{eq:gamma_phi_generalized} is valid for any constellation with equiprobable symbols. For constant-modulus modulation schemes, where $|\hat{X}|=|X|$ always hold, \eqref{eq:gamma_phi_generalized} simplifies to:
\begin{equation}
\Gamma_{\phi}(X,\hat{X})
= \sqrt{\frac{2}{N_0}}|h|\, \frac{-\Re\big\{X^{*}e^{-j\phi}(\hat{X}-X)\big\}}{\big|\hat{X}-X\big|}.
\label{eq:gamma_phi_special}
\end{equation}
In particular, for \ac{QPSK} with $X=\sqrt{E_\mathrm{s}}e^{j\theta_m}$, the neighboring symbols are $\hat{X}=\sqrt{E_\mathrm{s}}e^{j(\theta_m+\psi)}$, where $\psi$ is the angular separation; $\psi=\pi/2$ for the nearest neighbors. In this case:
\begin{equation}
    \big|\hat{X}-X\big|^{2}=2E_\mathrm{s}(1-\cos\psi),
\end{equation}
and the conditional \ac{PEP} becomes:
\begin{equation}
\mathbb{P}\big\{X\rightarrow\hat{X} \mid \phi,|h|\big\} = Q\!\left(
\sqrt{\gamma_{\mathrm{s}}}\,|h|\,
\frac{\cos\phi-\cos(\phi-\psi)}
{\sqrt{1-\cos\psi}}
\right),
\end{equation}
where $\gamma_{\mathrm{s}}=E_\mathrm{s}/N_0$ denotes the symbol \ac{SNR}. For nonconstant-modulus constellations including \ac{16QAM}, the conditional \ac{PEP} is averaged over the \ac{PDF} of the residual phase based on \eqref{eq:empirical_phase_pdf}:
\begin{equation}
\begin{aligned}
\mathbb{P}\big\{X\rightarrow\hat{X}\mid |h|\big\}
&= \mathbb{E}_{\phi}\Big\{
\mathbb{P}\big\{X\rightarrow\hat{X}\mid \phi,|h|\big\}
\Big\}, \\
&= \int_{-\infty}^{\infty}
Q\!\left(\Gamma_{\phi}(X,\hat{X})\right)
f_{\phi}(\phi)\,d\phi,
\end{aligned}
\label{eq:avgphase_PEP}
\end{equation}
where $\Gamma_{\phi}(X,\hat{X})$ is given by \eqref{eq:gamma_phi_generalized}. This integral does not have a closed-form solution, but it can be approximated using the Holtzman's three-point Gaussian method \cite{benedetto2002principles}:
\begin{equation}
\mathbb{E}_{\phi}\big\{g(\phi)\big\} \approx \frac{2}{3} g\big(\mu_{\phi}\big) + \frac{1}{6} g\big(\mu_{\phi}+\sqrt{3}\sigma_{\phi}\big) + \frac{1}{6} g\big(\mu_{\phi}-\sqrt{3}\sigma_{\phi}\big),
\label{11}
\end{equation}
through the use of $g(\phi) = Q\!\left(\Gamma_{\phi}(X,\hat{X})\right)$, where $\mu_{\phi}$ and $\sigma_{\phi}$ are the mean and standard deviation of $\phi$, respectively. The unconditional \ac{PEP} is obtained as:
\begin{equation}
\begin{aligned}
\mathbb{P}\big\{X\rightarrow\hat{X}\big\}
&= \mathbb{E}_{|h|} \Big\{\mathbb{P}\big\{X\rightarrow\hat{X}\mid |h|\big\}\Big\}, \\
&= \int_{0}^{\infty} \mathbb{P}\big\{X\rightarrow\hat{X}\mid u\big\}\, f_{|h|}(u)\,du,
\end{aligned}
\label{eq:PEP_fading_avg}
\end{equation}
where $f_{|h|}(u)$ is the \ac{PDF} of $|h|=G|h_\mathrm{turb}|$. The turbulence-induced fading magnitude $|h_\mathrm{turb}|=\sqrt{I_{\mathrm{r}}}$, where the irradiance $I_{\mathrm{r}}$ follows the Gamma-Gamma \ac{PDF} $f_{I_{\mathrm{r}}}(I_{\mathrm{r}})$ in \eqref{eq:gamma_gamma_pdf}. Using the transformation of random variables, it can be shown that:
\begin{equation}
    f_{|h|}(u) = \frac{2u}{G} f_{I_{\mathrm{r}}}\!\!\left(\frac{u^2}{G^2}\right).
\end{equation}
The integral in \eqref{eq:PEP_fading_avg} is evaluated numerically for each symbol pair using the Gauss-Laguerre quadrature. Furthermore, for an $M$-ary constellation with equiprobable symbols, the average \ac{SER} is upper-bounded by the union bound \cite{Proakis2008DigitalCommunications}:
\begin{equation}
P_{\mathrm{s}}
= \frac{1}{M}\sum_{X\in\mathcal{X}} \mathbb{P}\big\{\hat{X}\neq X\mid X\big\}
\leq \frac{1}{M}\sum_{X\in\mathcal{X}}
\sum_{\substack{\hat{X}\in\mathcal{X}\\ \hat{X}\neq X}}
\mathbb{P}\big\{X\rightarrow\hat{X}\big\}.
\label{eq:SER_union_general}
\end{equation}
For symmetric constellations such as square \ac{16QAM}, symbol pairs are classified into groups that share the same Euclidean distance $d_{\ell}=|\hat{X}-X|$ and average \ac{PEP} $\bar{P}_{\ell} = \mathbb{P}\big\{X\rightarrow\hat{X}\big\}$. In this case, \eqref{eq:SER_union_general} reduces to:
\begin{equation}
    P_{\mathrm{s}} \leq \sum_{\ell} N_{\ell}\bar{P}_{\ell},
\end{equation}
where $N_{\ell}$ is the average number of neighbors in the $\ell$th group.

\begin{table}[t!]
\centering
\caption{System Parameters}
\begin{tabular}{c|l|l}
\textbf{Symbol} & \textbf{Description} & \textbf{Value} \\
\hline
$\lambda$                 & Laser wavelength                    & $1550$ nm \\
--                        & Laser linewidth                     & $100$ kHz \\
$R_{\mathrm{s}}$          & Symbol rate                         & $40$ GBaud \\
$P_{\mathrm{t}}$          & Launch optical power                & $30$ dBm \\
$V_{\pi}$                 & Half-wave voltage                   & $4$ V \\
$V_{\mathrm{DC}}$         & DC bias voltage                     & $-2$ V \\
$\eta_{\mathrm{tx}}$      & Transmitter optical efficiency      & $0.6$ \\
$\eta_{\mathrm{rx}}$      & Receiver optical efficiency         & $0.6$ \\
$R_{\mathrm{PD}}$         & PD responsivity                     & $0.8$ A/W \\
$R_{\mathrm{E}}$          & Earth radius                        & $6357$ km \\
$\mu_{\mathrm{E}}$        & Geocentric gravitational constant   & $3.986\times10^{14}$ m$^3$/s$^2$ \\
$A_{\mathrm{HV}}$         & Hufnagel-Valley turbulence constant & $1.7\times10^{-14}$ m$^{-2/3}$ \\
$\sigma_\mathrm{I}^2$     & Scintillation index                 & $1.6$ \\
$\alpha_{\mathrm{GG}}$    & Gamma-Gamma PDF parameter           & $4$ \\
$\beta_{\mathrm{GG}}$     & Gamma-Gamma PDF parameter           & $1.9$ \\
$a$                       & Receiver aperture radius            & $15$ cm \\
$\sigma_{\mathrm{pe}}$    & RMS pointing jitter                 & $2$ $\mu$rad \\
$\gamma_{\mathrm{ho}}$    & DD-FLL handover margin              & $1.5$ \\
\hline
\end{tabular}
\label{tab:sys_param}
\end{table}

\section{Numerical Results and Discussions} \label{sec:results}
In this section, the proposed \ac{HTF-FOC} method for tracking and correction of \ac{CFO} in an \ac{LEO} satellite-to-ground coherent \ac{FSO} link is verified using computer simulations. Monte Carlo simulations are performed for a single-polarization downlink system with the end-to-end architecture shown in Fig.~\ref{fig:system}. The \ac{PEP} and \ac{SER} performance under the residual \ac{CFO} is studied based on the analysis from Section~\ref{sec:pep}. The system parameters used in the simulations are listed in Table~\ref{tab:sys_param}. In addition, the baseband waveform is oversampled at four samples per symbol before pulse shaping to reduce aliasing in the presence of \ac{ADC} sampling errors, assuming a sampling phase error of $0.10$ and a sampling frequency error of $\pm10$~ppm. The following results are presented under a moderate atmospheric turbulence based on the parameters given in Table~\ref{tab:sys_param}, for typical orbital altitudes of $H=400,600,800$~km and satellite velocities of $v_{\mathrm{sat}}=7.3,7.6,7.9$~km/s \cite{10138358,shoji2012pilot}.

\begin{figure*}[t!]
    \centering
    \begin{subfigure}[t]{0.24\textwidth}
        \centering
        \includegraphics[width=\linewidth]{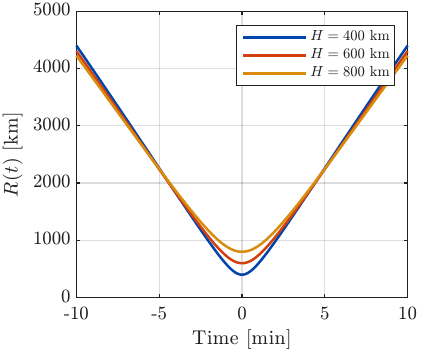}
        \caption{Slant range $R$ vs. Time}
    \end{subfigure}\hfill
    \begin{subfigure}[t]{0.24\textwidth}
        \centering
        \includegraphics[width=\linewidth]{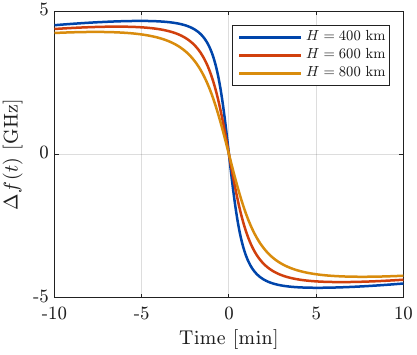}
        \caption{Doppler shift $\Delta f$ vs. Time}
    \end{subfigure}\hfill
    \begin{subfigure}[t]{0.24\textwidth}
        \centering
        \includegraphics[width=\linewidth]{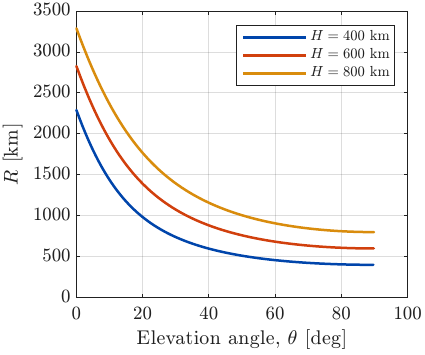}
        \caption{Slant range $R$ vs. Elevation angle $\theta$}
    \end{subfigure}\hfill
    \begin{subfigure}[t]{0.24\textwidth}
        \centering
        \includegraphics[width=\linewidth]{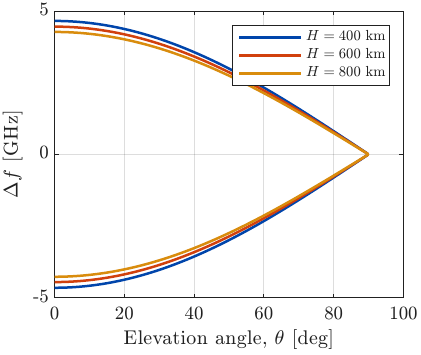}
        \caption{Doppler shift $\Delta f$ vs. Elevation angle $\theta$}
    \end{subfigure}
    \vspace{5pt}
    \begin{subfigure}[t]{0.24\textwidth}
        \centering
        \includegraphics[width=\linewidth]{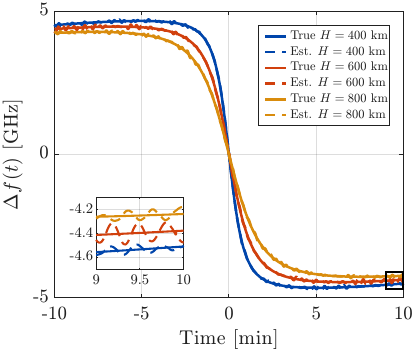}
        \caption{CFO tracking w.r.t. Altitude $H$}
    \end{subfigure}\hfill
    \begin{subfigure}[t]{0.24\textwidth}
        \centering
        \includegraphics[width=\linewidth]{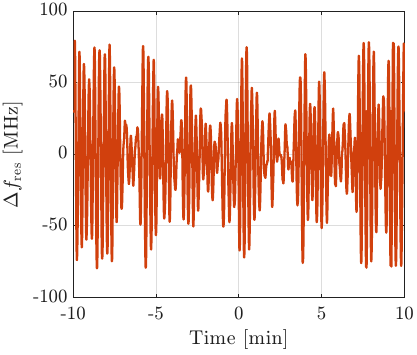}
        \caption{Residual CFO for $v_{\mathrm{sat}}=7.6$~km/s}
    \end{subfigure}\hfill
    \begin{subfigure}[t]{0.24\textwidth}
        \centering
        \includegraphics[width=\linewidth]{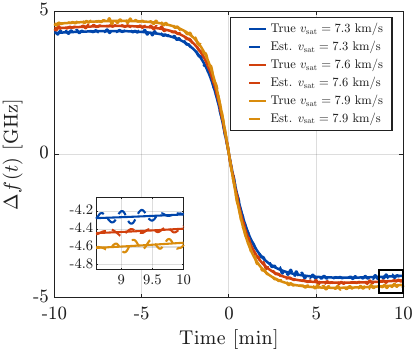}
        \caption{CFO tracking w.r.t. Orbital velocity $v_{\mathrm{sat}}$}
    \end{subfigure}\hfill
    \begin{subfigure}[t]{0.24\textwidth}
        \centering
        \includegraphics[width=\linewidth]{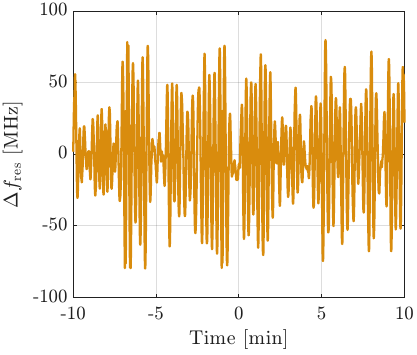}
        \caption{Residual CFO for $v_{\mathrm{sat}}=7.9$~km/s}
    \end{subfigure}
    \caption{Slant range, Doppler shift, and \ac{CFO} tracking for \ac{LEO} satellite passes with altitudes of $400\leq H\leq 800$~km and orbital speeds of $7.3\leq v_{\mathrm{sat}}\leq 7.9$~km/s. In (f), (h): $H=600$~km and $v_{\mathrm{sat}}=7.6$~km/s.\vspace{-5pt}}
    \label{fig:leo_cfo_8panels}
\end{figure*}

\subsection{CFO Tracking Performance}
Fig.~\ref{fig:leo_cfo_8panels} shows the geometric evolution of the \ac{LEO} satellite-to-ground pass and the corresponding \ac{CFO} tracking performance for three orbital altitudes and velocities using \ac{16QAM}. In Fig.~\ref{fig:leo_cfo_8panels}(a), the slant range $R(t)$ exhibits the expected V-shaped profile over a $20$~min pass, with the minimum range occurring near the time of closest approach. As the satellite altitude increases, the minimum range becomes larger, and the pass duration slightly increases. Fig.~\ref{fig:leo_cfo_8panels}(c) presents the same slant range as a function of the elevation angle, exhibiting a monotonic decrease of $R$ as the satellite rises from the horizon to the zenith, with higher orbits resulting in larger ranges at any given elevation angle. The associated Doppler shift is shown in Fig.~\ref{fig:leo_cfo_8panels}(b) as a function of time, where $\Delta f(t)$ starts at about $+5$~GHz when the satellite first becomes visible, then smoothly transitions through zero around the time of closest approach before settling near $-5$~GHz as the satellite recedes. Fig.~\ref{fig:leo_cfo_8panels}(d) plots the Doppler shift versus the elevation angle. The curve is almost symmetric with respect to the zenith. In Fig.~\ref{fig:leo_cfo_8panels}(e), the estimated \ac{CFO} closely follows the true GHz-level Doppler shifts for all three altitudes, including the steep transition region around the zenith. The residual frequency offset after tracking remains within $77.53$~MHz for $H=600$~km, as shown in Fig.~\ref{fig:leo_cfo_8panels}(f). When the orbital velocity varies as shown in Fig.~\ref{fig:leo_cfo_8panels}(g), the \ac{HTF-FOC} tracking loop accurately follows the true Doppler shifts for all cases of $v_{\mathrm{sat}}=7.3, 7.6, 7.9$~km/s, with a maximum deviation of approximately $79.81$~MHz for $v_{\mathrm{sat}}=7.9$~km/s, as shown in Fig.~\ref{fig:leo_cfo_8panels}(h). This confirms that the proposed \ac{CFO} tracking method performs well under different \ac{LEO} satellite altitudes and velocities.

\begin{figure}[t!]
    \centering
    \begin{subfigure}[t]{\linewidth}
        \centering
        \includegraphics[width=\linewidth]{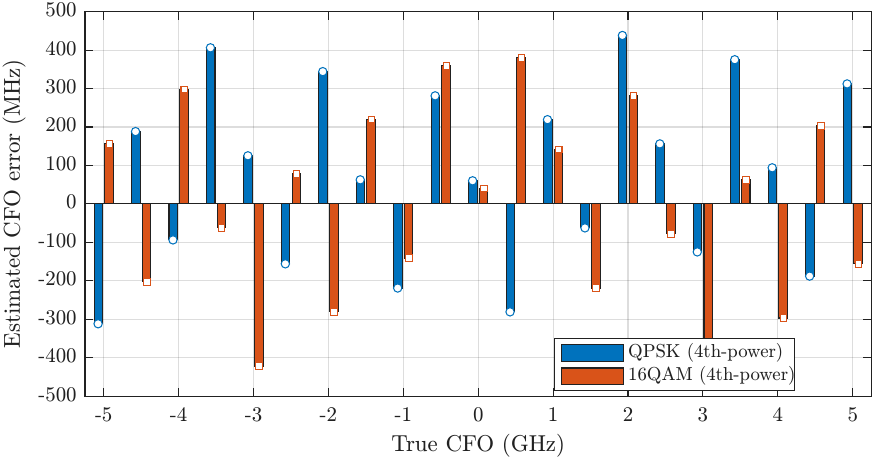}
        \caption{4th-power FFT-based coarse estimator}
    \end{subfigure}
    \vspace{5pt}
    \begin{subfigure}[t]{\linewidth}
        \centering
        \includegraphics[width=\linewidth]{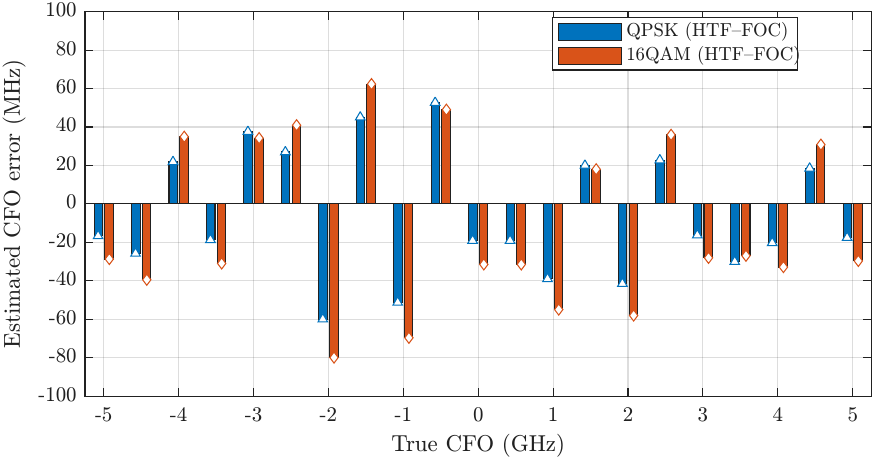}
        \caption{Proposed HTF-FOC tracking loop}
    \end{subfigure}
    \caption{CFO estimation error vs. True CFO for QPSK and 16QAM.\vspace{-10pt}}
    \label{fig:cfo_residual_htf_foc}
\end{figure}

Fig.~\ref{fig:cfo_residual_htf_foc} compares the estimated \ac{CFO} error obtained by the 4th-power FFT-based coarse acquisition with the proposed \ac{HTF-FOC} tracking loop for both \ac{QPSK} and \ac{16QAM} across a wide range of true \ac{CFO} values. In Fig.~\ref{fig:cfo_residual_htf_foc}(a), the 4th-power estimator exhibits large residual errors, which grow with the magnitude of the true \ac{CFO} and can exceed $400~\mathrm{MHz}$ in both cases. As shown in Fig.~\ref{fig:cfo_residual_htf_foc}(b), when the \ac{HTF-FOC} tracking loop is enabled, the residual \ac{CFO} estimation error is significantly reduced. In this case, the residual error stays within approximately $\pm 80~\mathrm{MHz}$ across the entire $\pm 5~\mathrm{GHz}$ range of the true \ac{CFO} for both \ac{QPSK} and \ac{16QAM}, demonstrating that the proposed \ac{HTF-FOC} method effectively reduces the \ac{CFO}.

\begin{figure}[t!]
    \centering
    \begin{subfigure}[t]{0.32\linewidth}
        \centering
        \includegraphics[width=\linewidth]{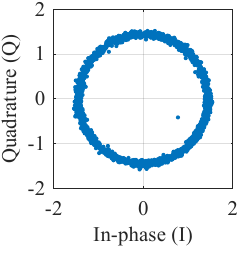}
        \caption{QPSK before FOC}
    \end{subfigure}\hfill
    \begin{subfigure}[t]{0.32\linewidth}
        \centering
        \includegraphics[width=\linewidth]{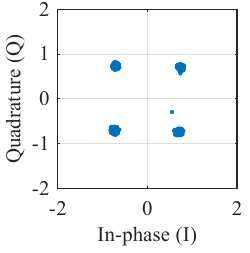}
        \caption{QPSK after HTF-FOC}
    \end{subfigure}\hfill
    \begin{subfigure}[t]{0.32\linewidth}
        \centering
        \includegraphics[width=\linewidth]{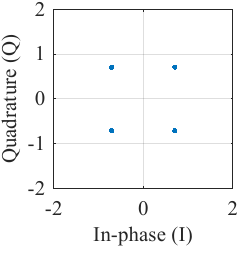}
        \caption{QPSK after CPR}
    \end{subfigure} \\ \vspace{5pt}
    \begin{subfigure}[t]{0.32\linewidth}
        \centering
        \includegraphics[width=\linewidth]{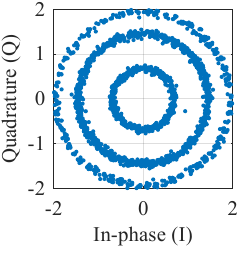}
        \caption{16QAM before FOC}
    \end{subfigure}\hfill
    \begin{subfigure}[t]{0.32\linewidth}
        \centering
        \includegraphics[width=\linewidth]{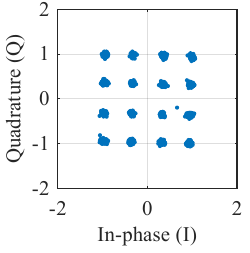}
        \caption{16QAM after HTF-FOC}
    \end{subfigure}\hfill
    \begin{subfigure}[t]{0.32\linewidth}
        \centering
        \includegraphics[width=\linewidth]{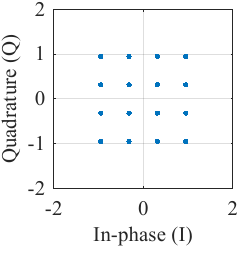}
        \caption{16QAM after CPR}
    \end{subfigure}
    \caption{CFO-induced constellation distortion, compensation by the HTF-FOC method, and phase recovery for QPSK and 16QAM, under $\Delta f=1$~GHz for $H=600$~km and $v_{\mathrm{sat}}=7.6$~km/s.\vspace{-5pt}}
    \label{fig:cfo_constellations}
\end{figure}

The presence of \ac{CFO} causes a continuous phase drift in the received signal, leading to a rotation of the signal constellation. For \ac{QPSK}, all constellation points rotate uniformly about the origin, as illustrated in Fig.~\ref{fig:cfo_constellations}. Despite a relatively large symbol spacing, the cumulative phase drift eventually results in decision errors. Fig.~\ref{fig:cfo_constellations}(a) shows a continuous circular drift caused by the uncompensated \ac{CFO}, while Fig.~\ref{fig:cfo_constellations}(b) and Fig.~\ref{fig:cfo_constellations}(c) show the constellation recovering to its ideal locations after \ac{CFO} compensation. For \ac{16QAM}, the impact is worse, as the constellation points rotate and spread out, and neighboring clusters begin to overlap, as shown in Fig.~\ref{fig:cfo_constellations}(d). By applying the \ac{HTF-FOC} algorithm, the signal constellation returns to clearly separated clusters, with a small residual jitter near the decision boundaries, as shown in Fig.~\ref{fig:cfo_constellations}(e) and Fig.~\ref{fig:cfo_constellations}(f).

\subsection{PEP and SER Performance}
\begin{figure}[t!]
    \centering
    \begin{subfigure}[t]{\linewidth}
        \centering
        \includegraphics[width=\columnwidth]{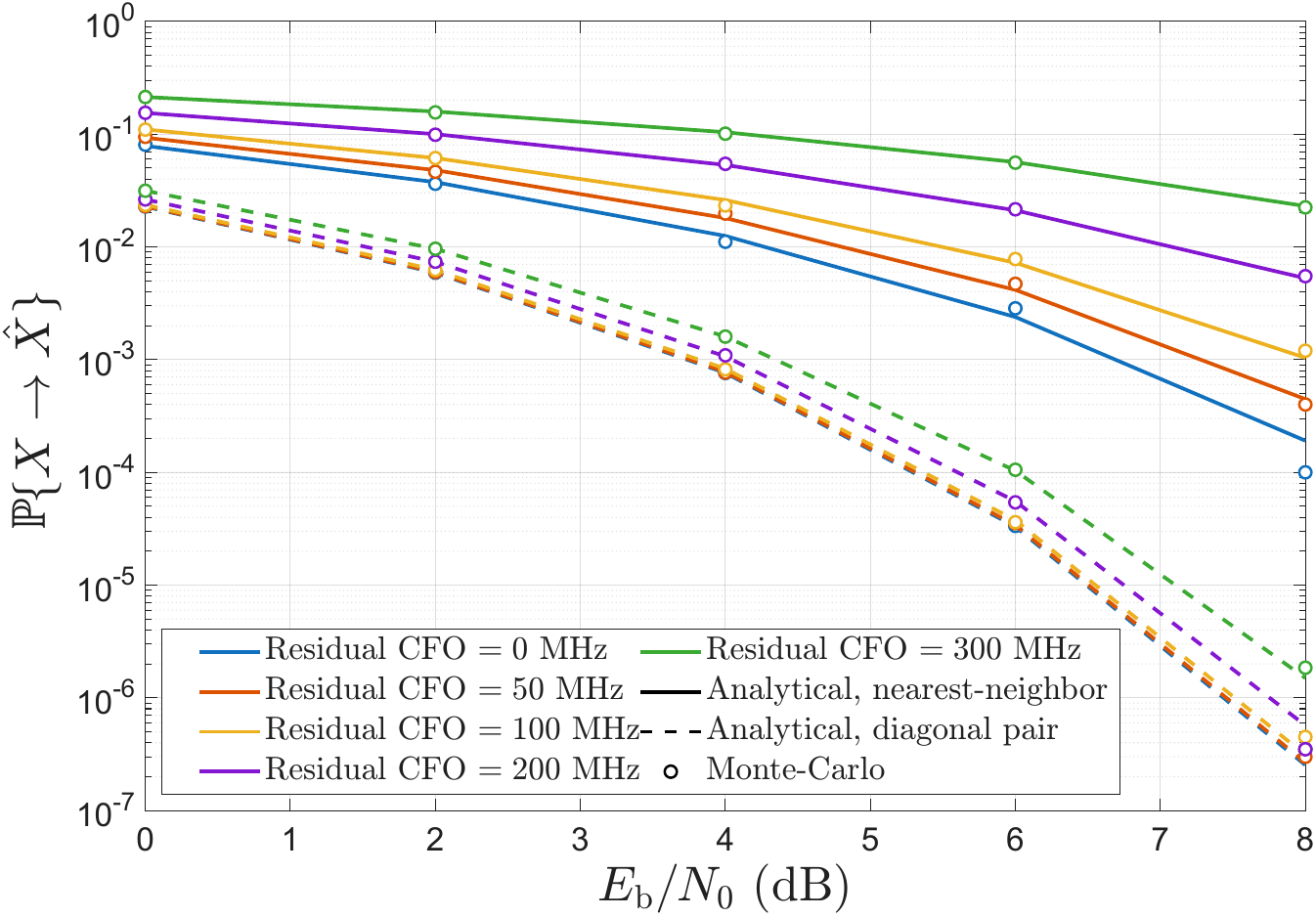}
        \caption{QPSK using $K_\mathrm{p} = 0.05$ and $K_\mathrm{i} = 0.7$\vspace{5pt}}
    \end{subfigure}
    \begin{subfigure}[t]{\linewidth}
        \centering
        \includegraphics[width=\linewidth]{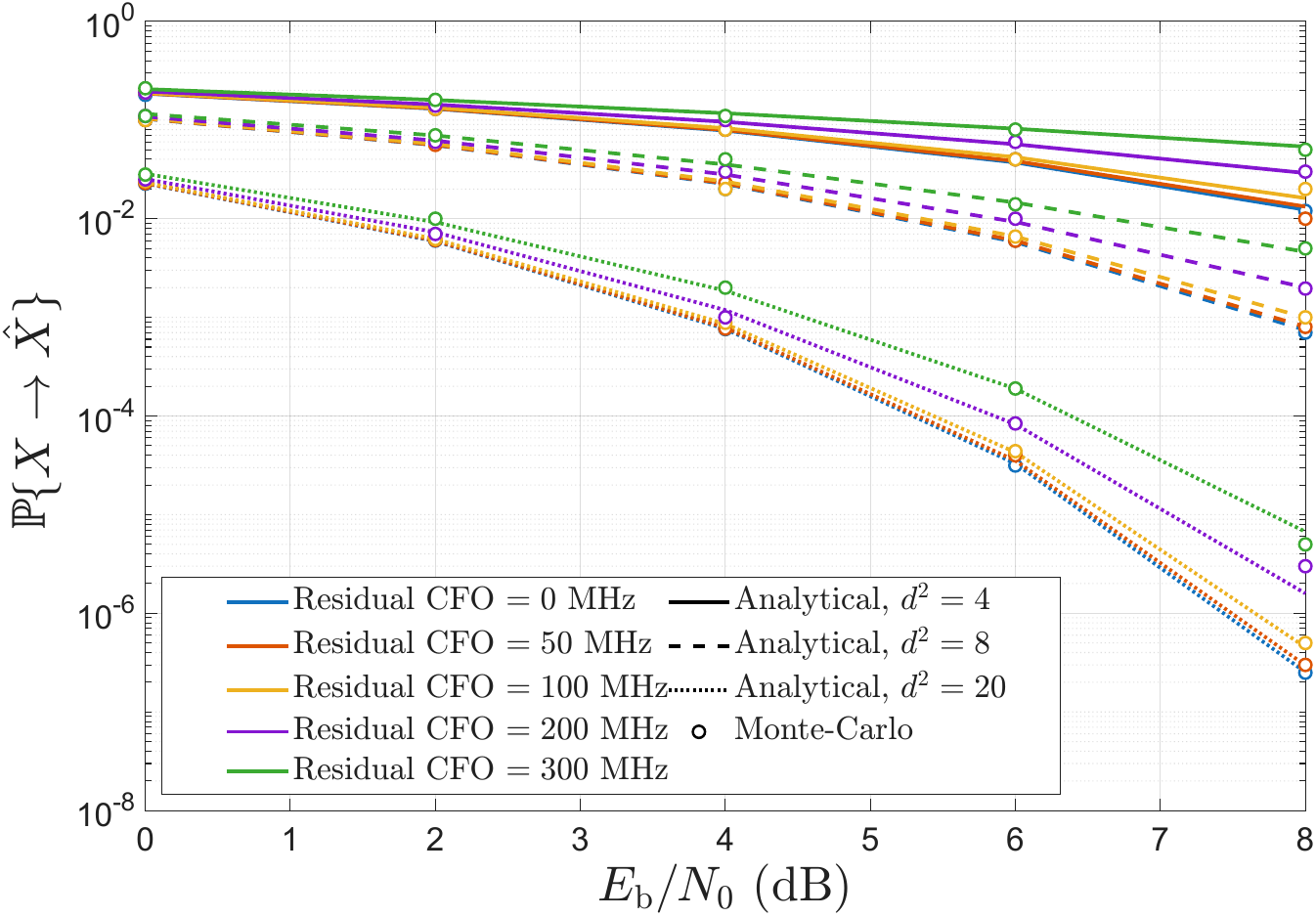}
        \caption{16QAM using $K_\mathrm{p} = 0.03$ and $K_\mathrm{i} = 0.5$}
    \end{subfigure}
    \caption{%
        PEP against $E_b/N_0$ for HTF-FOC with different post-compensation residual \ac{CFO} values.\vspace{-10pt}}
    \label{fig:pep_qpsk_16qam_cfo}
\end{figure}

Fig.~\ref{fig:pep_qpsk_16qam_cfo}(a) demonstrates the impact of the post-compensation residual \ac{CFO} on the \ac{PEP} performance for \ac{QPSK}. Observing the estimated \ac{CFO} error bound of $\pm80$~MHz over the full \ac{CFO} range of $\pm5$~GHz after \ac{HTF-FOC}, the question is how the sensitivity of the tracking loop changes for residual \ac{CFO} values beyond this limit. In Fig.~\ref{fig:pep_qpsk_16qam_cfo}(a), five operating points from $0$ to $300$~MHz are specified for the residual \ac{CFO}. For each case, the corresponding discrete-time residual phase sequence is generated and used to evaluate the average \ac{PEP} analytically and in the Monte Carlo domain, as discussed in Section~\ref{sec:pep}. It can be seen that the analytical and Monte Carlo simulation results are closely matched for different values of the residual \ac{CFO}. Both Euclidean distance classes in \ac{QPSK}, namely the nearest-neighbors $(d^2 = 4)$ and diagonal pairs $(d^2 = 8)$, exhibit only a minor performance degradation when the residual \ac{CFO} is small. For example, for the diagonal class at $E_b/N_0 = 8$~dB, the \ac{PEP} increases only slightly from $2.53\times10^{-7}$ at $\Delta f = 0$ to $3.11\times10^{-7}$ at $\Delta f = 100$~MHz. However, at $\Delta f=300$~MHz, the \ac{PEP} for the diagonal pair increases to $1.85\times10^{-6}$, and for the nearest neighbor reaches $2.4\times10^{-2}$ at the same $E_\text{b}/N_\text{0}$. These results confirm that small residual \ac{CFO} values below $100$~MHz have a marginal effect on the \ac{PEP} performance, while larger \ac{CFO} values significantly degrade the carrier frequency tracking performance.

Fig.~\ref{fig:pep_qpsk_16qam_cfo}(b) presents the \ac{PEP} performance for different residual \acp{CFO} for \ac{16QAM}, with a rectangular constellation using \ac{IQ} levels of $\{\pm1,\pm3\}$. The pairwise error events fall within three distinct Euclidean distance classes. The first class includes the nearest neighbors along the horizontal or vertical axis, which differ by $\pm2$ on one axis and $0$ on the other, leading to $d^{2}=4$. The second class corresponds to the diagonal neighbors, where two symbols differ by $\pm2$ on both axes, which means $d^{2}=8$. The third class involves a longer transition with a difference of $4$ on one axis and $2$ on the other, resulting in $d^{2}=20$. It is evident that the analytical \ac{PEP} results are in close agreement with the Monte Carlo simulations for different values of the residual \ac{CFO}. The nearest-neighbor class is the most impactful case in terms of the \ac{PEP} performance due to its smallest Euclidean distance. For $d^{2}=20$ and $E_b/N_0 = 8$~dB, the \ac{PEP} increases only slightly from $2.53\times10^{-7}$ at $\Delta f = 0$ to $4.53\times10^{-7}$ at $\Delta f=100$~MHz. By contrast, a large \ac{CFO} of $\Delta f=300$~MHz increases the \ac{PEP} to $6.72\times10^{-6}$. This indicates that small \ac{CFO} values have an insignificant effect on the symbol decision performance, whereas large \ac{CFO} values lead to considerable \ac{PEP} penalties.

\begin{figure}[t!]
    \centering
    \includegraphics[width=\columnwidth]{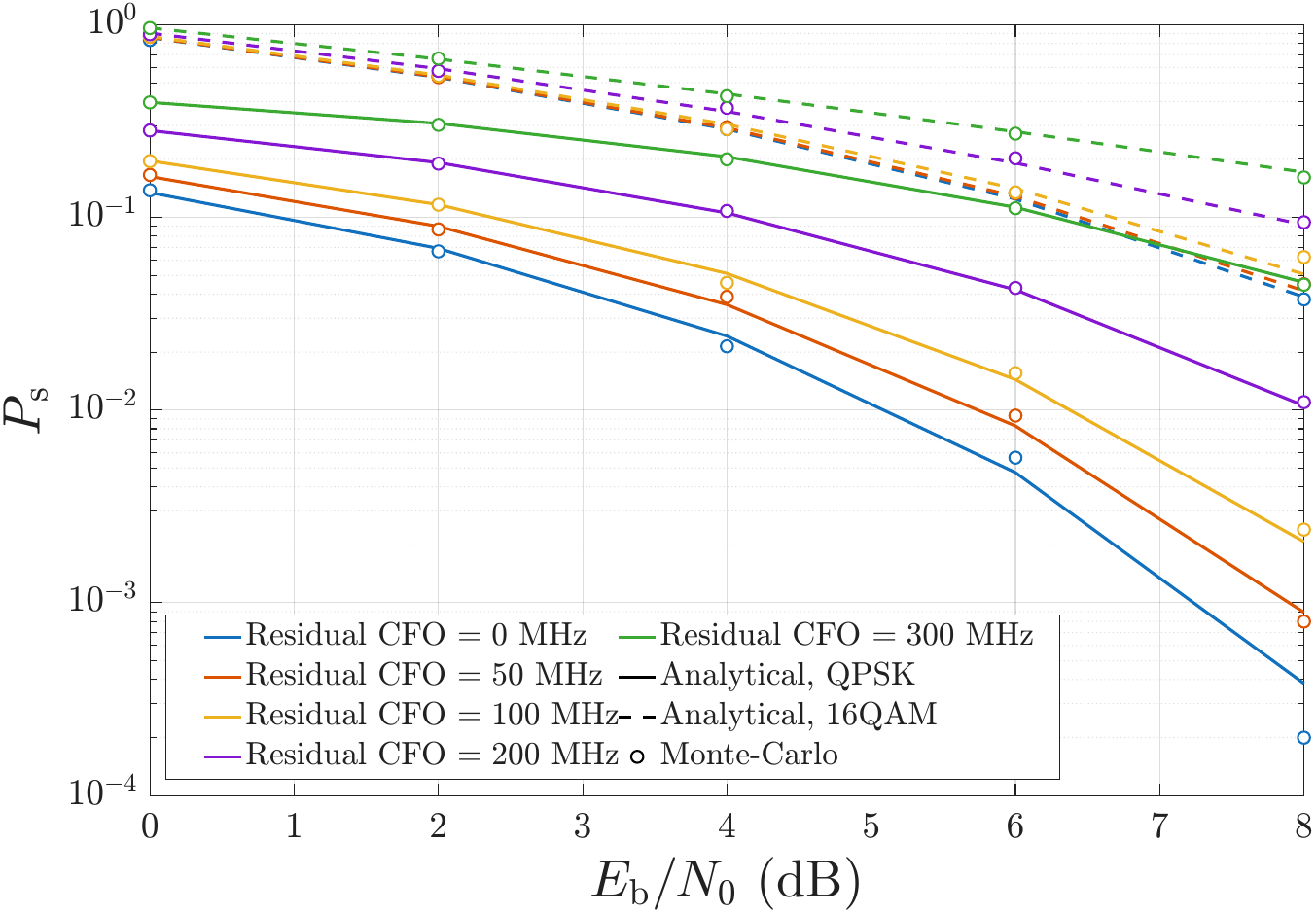}
    \caption{SER $P_{\mathrm{s}}$ against $E_{\mathrm{b}}/N_{\mathrm{0}}$ for QPSK and 16QAM under different post-compensation residual CFO values using HTF-FOC, with $K_\mathrm{p} = 0.05$ and $K_\mathrm{i} = 0.7$ for QPSK, $K_\mathrm{p} = 0.03$ and $K_\mathrm{i} = 0.5$ for 16QAM.\vspace{-5pt}}
    \label{fig:ser_qpsk_16qam_cfo}
\end{figure}

Fig. ~\ref{fig:ser_qpsk_16qam_cfo} illustrates the \ac{SER} performance as a function of $E_{\mathrm{b}}/N_{\mathrm{0}}$ for different values of the post-compensation residual \ac{CFO} for both \ac{QPSK} and \ac{16QAM}. In both cases, the analytical and Monte Carlo simulation results are closely matching. As the residual \ac{CFO} increases, the \ac{SER} curves are shifted upward, indicating the performance loss caused by imperfect residual frequency offset compensation. For a given value of $E_{\mathrm{b}}/N_{\mathrm{0}}$ and \ac{CFO}, \ac{16QAM} has a higher \ac{SER} than \ac{QPSK} because of its smaller minimum Euclidean distance, which is consistent with the reliability-spectral efficiency trade-off. In both cases, small \ac{CFO} values of $\Delta f < 100$~MHz have a negligible effect on the \ac{SER} performance. For example, for $E_b/N_0 = 8$~dB, the \ac{SER} increases from $P_{\mathrm{s}}=3.81\times10^{-4}$ at $\Delta f = 0$ to $P_{\mathrm{s}}=2.07\times10^{-3}$ at $\Delta f = 100$~MHz. By comparison, a large \ac{CFO} of $\Delta f = 300$~MHz raises the \ac{SER} to $P_{\mathrm{s}}=4.4\times10^{-2}$, which is a noticeable degradation. Since the proposed \ac{HTF-FOC} method keeps the residual \ac{CFO} below $80$~MHz for different altitudes and velocities, it ensures reliable demodulation.

\subsection{Comparison with Baseline FOC Methods}
To gain a deeper understanding of the performance, Table~\ref{tab:baseline_evm_lock_qpsk_16qam} compares the proposed \ac{HTF-FOC} with a number of practical coarse-to-fine \acs{FOC} baseline techniques in terms of \ac{EVM} penalty and lock rate, defined as the ratio of successful to total number of \ac{CFO} acquisition trials, assuming the same received signal and moderate turbulence. The \ac{VV} method uses a short 4th-power phase increment acquisition followed by a \ac{DD}-\ac{FLL} \cite{viterbi1983nonlinear}. The pilot \ac{CFO} method uses a known symbol sequence of length $32$ for initial \ac{CFO} estimation, then applies a \ac{DD}-\ac{FLL} \cite{zhao2015digital}. The \ac{FFT}+\ac{DD}-Kalman method adopts the \ac{FFT}-based coarse \ac{CFO} acquisition and performs fine \ac{CFO} tracking using a first order residual frequency state model \cite{xu2025robust}. The proposed method offers the lowest \ac{EVM} penalty and maintains the lock rate at $100\%$ for both \ac{QPSK} and \ac{16QAM}. To verify this performance under different turbulence conditions, the handover ratio is defined as:
\begin{equation}
\mathcal{R}_{\mathrm{ho}} = \frac{|\hat{\omega}_{\mathrm{res},0}|}{\epsilon_{\mathrm{lock}}\omega_{\max}},
\end{equation}
where $\mathcal{R}_{\mathrm{h}}<1$ is the stability condition in Algorithm~\ref{alg:stability_check}. Based on the weak, moderate, and strong Gamma-Gamma turbulence parameters from Fig.~\ref{fig:cn2_gamma}, the proposed method successfully completes all $50$ trials, with corresponding handover ratios of $\mathcal{R}_{\mathrm{ho}}=0.02,0.01,0.01$ for \ac{QPSK}, and $\mathcal{R}_{\mathrm{ho}}=0.05,0.07,0.07$ for \ac{16QAM}, confirming that it has robust performance under any turbulence channel condition.

\begin{table}[t!]
\centering
\caption{Mean EVM penalty and carrier-acquisition success rate under time-varying Doppler over 50 Monte Carlo trials.}
\label{tab:baseline_evm_lock_qpsk_16qam}
\setlength{\tabcolsep}{3pt}
\begin{tabular}{l|cc|cc}
 & \multicolumn{2}{c}{\textbf{EVM Penalty (dB)}} & \multicolumn{2}{c}{\textbf{Lock Rate (\%)}} \\
 \hline
\textbf{Method} & QPSK & 16QAM & QPSK & 16QAM \\
\hline
Proposed HTF-FOC & 4.20 & 10.35 & 100 & 100 \\
VV coarse \cite{viterbi1983nonlinear} + DD-FLL & 14.36 & 17.26 & 100 & 46 \\
Pilot CFO \cite{zhao2015digital} + DD-FLL & 16.20 & 17.49 & 100 & 92 \\
FFT + DD-Kalman \cite{xu2025robust} & 17.68 & 17.98 & 96 & 44 \\
DD-FLL only & 18.06 & 18.07 & 0 & 0 \\
\hline
\end{tabular}
\end{table}

\section{Conclusions} \label{sec:conclusions}
This paper investigated Doppler-induced \ac{CFO} estimation and tracking in \ac{LEO} satellite-to-ground coherent \ac{FSO} links, where orbital motion produces time-varying multi-GHz frequency offsets. To overcome the limited capture range of conventional carrier recovery loops and enable wideband \ac{CFO} acquisition based on \ac{DSP}, a novel Doppler-aware \ac{HTF-FOC} method was proposed and developed, encompassing residual coarse \ac{CFO} estimation and handover followed by per-symbol \ac{CFO} tracking using the \ac{DD}-\ac{FLL}. A handover checking mechanism was devised to activate the \ac{DD}-\ac{FLL}, via approximating the \ac{FFT}-based \ac{CFO} acquisition by a short estimation window with an alias-free range of $|\Delta f|<F_{\mathrm{s}}/8$, with $F_{\mathrm{s}}$ representing the sampling frequency. The results evince that the proposed \ac{HTF-FOC} method can track Doppler-induced \ac{CFO} variations exceeding $\pm5$~GHz for \ac{LEO} altitudes of $400\leq H\leq 800$~km and orbital velocities of $7.3\leq v_{\mathrm{sat}}\leq 7.9$~km/s. It significantly improves the residual \ac{CFO} performance from above $400$~MHz provided by the 4th-power coarse estimator to below $80$~MHz after fine tracking by the \ac{DD}-{FLL}. The baseline \ac{VV} and pilot \ac{CFO} coarse estimation methods are unable to acquire $\pm5$~GHz \acp{CFO}, and the \ac{FFT}+\ac{DD}-Kalman method provides lock rates of $96\%$ and $44\%$ for \ac{QPSK} and \ac{16QAM}, respectively, while the proposed method attains the $100\%$ lock rate for both cases. However, for the \ac{FFT}+\ac{DD}-Kalman method, the \ac{EVM} penalty exceeds $18$~dB, while the proposed method brings this down to $4.20$~dB and $10.35$~dB for \ac{QPSK} and \ac{16QAM}, respectively. In addition, the proposed method provides a robust performance by keeping handover ratios under $0.02$ for \ac{QPSK} and $0.07$ for \ac{16QAM} regardless of the turbulence-induced fading strength. The average \ac{PEP} and \ac{SER} were derived to study the impact of residual \acp{CFO} beyond $80$~MHz. The results show that, when the residual \ac{CFO} $\leq100$~MHz, the average \ac{SER} remains below $2.1\times10^{-3}$ for $E_b/N_0\geq8$~dB. The proposed Doppler-aware \ac{CFO} acquisition supports integrated carrier recovery and resource allocation in future satellite coherent \ac{FSO} networks.

\bibliographystyle{IEEEtran}  
\bibliography{references_v10}

\begin{IEEEbiography}[{\includegraphics[width=1in,height=1.25in,clip,keepaspectratio]{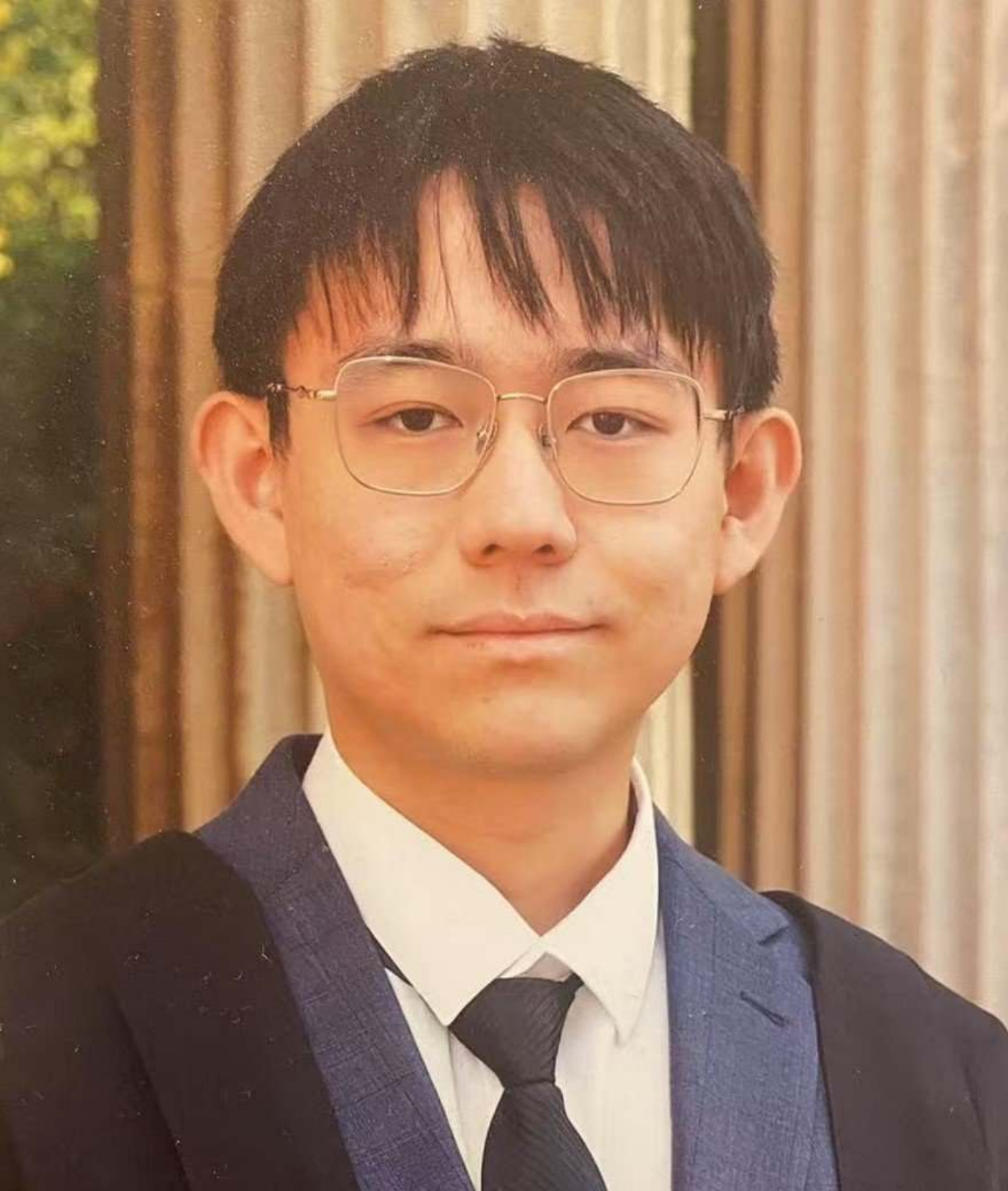}}]{Tiankuo Jiao}
received the B.Eng. degree in Electrical and Electronic Engineering with First Class Honors from the University of Liverpool, U.K., in 2023. He subsequently earned an M.Res. degree in Photonic and Electronic Systems from the University of Cambridge, U.K., in 2024. He is currently pursuing a Ph.D. at the LiFi Research and Development Center (LRDC), University of Cambridge, where his research focuses on the design and development of optical wireless integrated sensing and communication (OW-ISAC) systems and advanced digital signal processing (DSP) algorithms for coherent free-space optical (FSO) communications.
\end{IEEEbiography}


\begin{IEEEbiography}[{\includegraphics[width=1in,height=1.25in,clip,keepaspectratio]{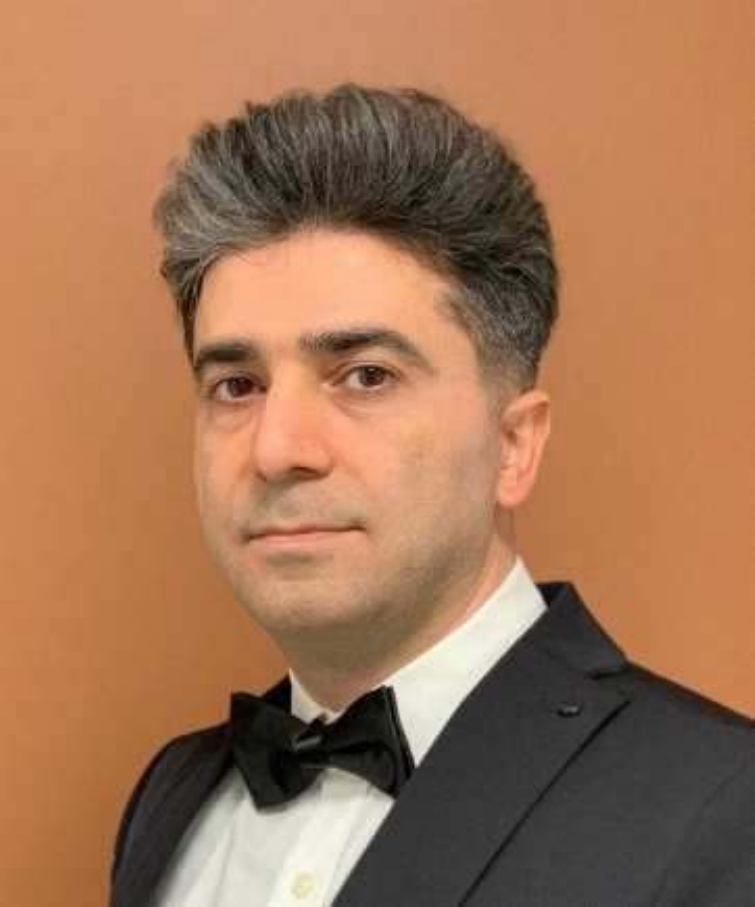}}]{Hossein Kazemi}
(Member, IEEE) received the Ph.D. degree in Electrical Engineering from The University of Edinburgh, U.K., in 2019. He also received the M.Sc. degree in Electrical Engineering (Microelectronic Circuits) from Sharif University of Technology, Tehran, Iran, in 2011, and the M.Sc. degree (Hons.) in Electrical Engineering (Wireless Communications) from Ozyegin University, Istanbul, Turkey, in 2014. He is a Postdoctoral Research Associate at the LiFi Research and Development Center, University of Cambridge, U.K. Dr Kazemi was the recipient of the Best Paper Award for the 2022 IEEE Global Communications Conference (GLOBECOM). His current research interests include the design, analysis and optimization of ultra-high-speed optical wireless communication systems for 6G and beyond networks.
\end{IEEEbiography}


\begin{IEEEbiography}[{\includegraphics[width=1in,height=1.25in,clip,keepaspectratio]{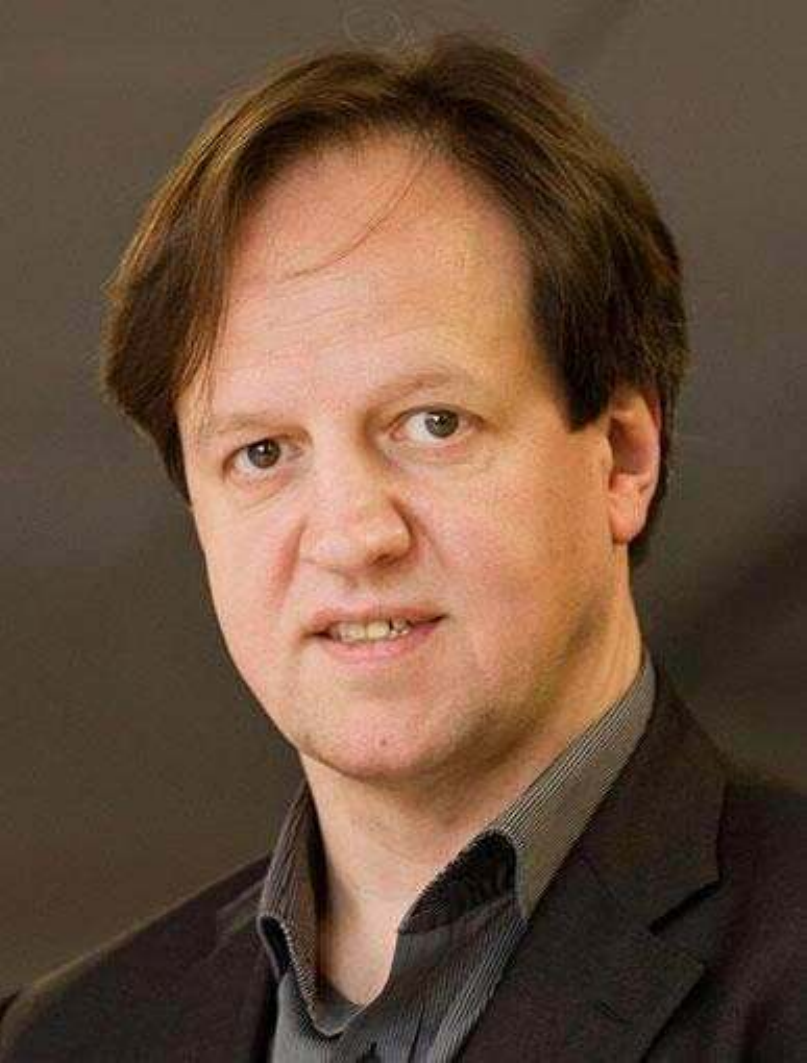}}]{Harald Haas} 
(Fellow, IEEE) received his Ph.D. from The University of Edinburgh, U.K., in 2001. He is the Van Eck Chair of Engineering at the University of Cambridge and the founder of pureLiFi Ltd., where he also serves as the Chief Scientific Officer (CSO). His recent research interests focus on photonics, communication theory and signal processing for optical wireless communication systems. Since 2017, he has been recognised as a highly cited researcher by Clarivate/Web of Science. He has delivered two TED talks and one TEDx talk. In 2016, he received the Outstanding Achievement Award from the International Solid State Lighting Alliance. He was awarded the Royal Society Wolfson Research Merit Award in 2017, the IEEE Vehicular Technology Society James Evans Avant Garde Award in 2019, and the Enginuity: The Connect Places Innovation Award in 2021. In 2022, he received the Humboldt Research Award for his research contributions. He is a Fellow of the Royal Academy of Engineering (RAEng), the Royal Society of Edinburgh (RSE), and the Institution of Engineering and Technology (IET). In 2023, he was shortlisted for the European Inventor Award.
\end{IEEEbiography}

\end{document}